\documentclass[12pt,a4]{article}

\usepackage{amsthm,amsmath,latexsym,amssymb,amsfonts,amscd}
\usepackage{graphics,lscape,fancyhdr,array,stmaryrd,euscript}
\usepackage{empheq}
\usepackage{dsfont}
\usepackage{verbatim}
\usepackage{color,tikz,tikz-cd}
\usetikzlibrary[snakes]
\usepackage{relsize,slashed}
\numberwithin{equation}{section}
\usepackage{hyperref}

\newcommand{\bep}{\begin{picture}}
\newcommand{\eep}{\end{picture}}

\newcommand{\smallpic}[1]{{\unitlength=0.2mm#1}}

\newcounter{YoungHeight}\newcounter{YoungWidth}

\newcounter{Mul1}\newcounter{Mul2}\newcounter{Mul3}\newcounter{Mul4}
\newcounter{A0}\newcounter{A1}\newcounter{A2}
\newcounter{B3}
\newcounter{C3}\newcounter{C4}
\newcounter{D1}\newcounter{D2}\newcounter{D3}
\newcounter{T0}\newcounter{T1}

\newlength{\txtHShift}

\newlength{\txtWidth}

\newcommand{\Add}[3]{\setcounter{#1}{#2}\addtocounter{#1}{#3}}

\newcommand{\Length}[1]{#10}
\newcommand{\YoungScale}{}

\newcommand{\BlockA}[2]{{\YoungScale\bep(\Length{#1},\Length{#2}){\Add{A1}{#1}{1}\Add{A2}{#2}{1}}%
\multiput(0,0)(10,0){\value{A1}}{\line(0,1){\Length{#2}}}\multiput(0,0)(0,10){\value{A2}}{\line(1,0){\Length{#1}}}%
\setcounter{YoungHeight}{\Length{#2}}\setcounter{YoungWidth}{\Length{#1}}\eep}}

\newcommand{\YoungB}{\BlockA{2}{1}}

\newcommand{\BlockApar}[2]{\parbox{\Length{#1}pt}{\YoungScale\bep(\Length{#1},\Length{#2}){\Add{A1}{#1}{1}\Add{A2}{#2}{1}}%
\multiput(0,0)(10,0){\value{A1}}{\line(0,1){\Length{#2}}}\multiput(0,0)(0,10){\value{A2}}{\line(1,0){\Length{#1}}}%
\setcounter{YoungHeight}{\Length{#2}}\setcounter{YoungWidth}{\Length{#1}}\eep}}

\newcommand{\BlockBpar}[4]{\parbox{\Length{#1}pt}{\YoungScale\Add{B3}{\Length{#2}}{\Length{#4}}%
\bep(\Length{#1},\value{B3})\put(0,\Length{#4}){\BlockA{#1}{#2}}%
\put(0,0){\BlockA{#3}{#4}}\setcounter{YoungHeight}{\value{B3}}\setcounter{YoungWidth}{\Length{#1}}\eep}}

\newcommand{\YoungpB}{\BlockApar{2}{1}}

\newcommand{\YoungpAA}{\BlockApar{1}{2}}

\newcommand{\YoungpBB}{\BlockApar{2}{2}}
\newcommand{\YoungpCA}{\BlockBpar{3}{1}{1}{1}}

\newcommand{\YoungpCC}{\BlockApar{3}{2}}

\newcommand{\YoungpAAAA}{\BlockApar{1}{4}}

\newcommand{\YoungpBBAA}{\BlockBpar{2}{2}{1}{2}}

\newcommand{\BlockAm}[2]{\mbox{\smallpic{\YoungScale\bep(\Length{#1},\Length{#2}){\Add{A1}{#1}{1}\Add{A2}{#2}{1}}%
\multiput(0,0)(10,0){\value{A1}}{\line(0,1){\Length{#2}}}\multiput(0,0)(0,10){\value{A2}}{\line(1,0){\Length{#1}}}%
\setcounter{YoungHeight}{\Length{#2}}\setcounter{YoungWidth}{\Length{#1}}\eep}}}

\newcommand{\BlockBm}[4]{\mbox{\smallpic{\YoungScale\Add{B3}{\Length{#2}}{\Length{#4}}%
\bep(\Length{#1},\value{B3})\put(0,\Length{#4}){\BlockA{#1}{#2}}%
\put(0,0){\BlockA{#3}{#4}}\setcounter{YoungHeight}{\value{B3}}\setcounter{YoungWidth}{\Length{#1}}\eep}}}

\newcommand{\YoungmA}{\BlockAm{1}{1}}
\newcommand{\YoungmB}{\BlockAm{2}{1}}
\newcommand{\YoungmC}{\BlockAm{3}{1}}

\newcommand{\YoungmAA}{\raisebox{-2pt}{\BlockAm{1}{2}}}
\newcommand{\YoungmBA}{\raisebox{-2pt}{\BlockBm{2}{1}{1}{1}}}
\newcommand{\YoungmBB}{\raisebox{-2pt}{\BlockAm{2}{2}}}

\newcommand{\YoungmAAAA}{\raisebox{-8pt}{\BlockAm{1}{4}}}
\newcommand{\YoungmAAA}{\raisebox{-4pt}{\BlockAm{1}{3}}}

\newcommand{\BlockSLA}[2]{\parbox{\Length{#1}pt}{\YoungScale\bep(\Length{#1},\Length{#2}){\linethickness{2mm}\Add{A1}{#1}{1}\Add{A2}{#2}{1}}%
\multiput(0,0)(10,0){\value{A1}}{\linethickness{0.4mm}\line(0,1){\Length{#2}}}\multiput(0,0)(0,10){\value{A2}}{\linethickness{0.4mm}\line(1,0){\Length{#1}}}%
\multiput(0,0)(10,0){#1}{\multiput(0,0)(0,10){#2}{\put(0,0){\line(1,1){10}}\put(0,\Length{1}){\line(1,-1){10}}}}\eep}}

\newcommand{\YoungSLAA}{\BlockSLA{1}{2}}

\newcommand{\YoungSLBB}{\BlockSLA{2}{2}}

\newcommand{\YoungSLCC}{\BlockSLA{3}{2}}

\newcommand{\pl}{\partial}

\usepackage{hyperref}

\usepackage[numbers,sort&compress]{natbib}
\usepackage[nottoc]{tocbibind}

\newcommand{\aA}{{\ensuremath{\mathcal{A}}}}
\newcommand{\aB}{{\ensuremath{\mathcal{B}}}}
\newcommand{\aC}{{\ensuremath{\mathcal{C}}}}
\newcommand{\aD}{{\ensuremath{\mathcal{D}}}}

\newcommand{\ga}{\alpha}
\newcommand{\gb}{\beta}
\newcommand{\gc}{\gamma}
\newcommand{\gd}{\delta}
\newcommand{\gad}{{\dot{\alpha}}}
\newcommand{\gbd}{{\dot{\beta}}}

\newcommand{\bry}{{{\bar{y}}}}

\newcommand{\brq}{{{\bar{q}}}}
\newcommand{\brk}{{{\bar{k}}}}

\newcommand{\fud}[2]{{}^{#1}{}_{#2}\,}
\newcommand{\fdu}[2]{{}_{#1}{}^{#2}\,}

\newcommand{\Pd}{{\mathtt{P}}}
\newcommand{\Ld}{{\mathtt{L}}}
\newcommand{\brLd}{\bar{\mathtt{L}}}

\newcommand{\Imap}{{\mathtt{I}}}

\newcommand{\besubeqs}{\begin{subequations}}
\newcommand{\esubeqs}{\end{subequations}}

\newcommand{\ass}{{\mathrm{A}}}
\newcommand{\Jos}{{\mathcal{J}}}

\newcommand{\orderA}[1]{\begin{tikzpicture}[scale=0.4]
\tikzset{%
  >=latex, 
  inner sep=0pt,%
  outer sep=2pt,%
  mark coordinate/.style={inner sep=0pt,outer sep=0pt,minimum size=3pt}%
}
\def\R{0.15}
\draw[black,thick] (1,1) -- (0,0);
\draw[black,thick] (1,1) -- (2,0);
\filldraw[color=black,fill=green]   (1,1)  circle (\R) node[left,xshift=-0.2cm]{$\scriptstyle #1$};
\end{tikzpicture}}
\newcommand{\orderB}[2]{\begin{tikzpicture}[scale=0.4]
\tikzset{%
  >=latex, 
  inner sep=0pt,%
  outer sep=2pt,%
  mark coordinate/.style={inner sep=0pt,outer sep=0pt,minimum size=3pt}%
}
\def\R{0.15}
\draw[black,thick] (1,1) -- (0,0);
\draw[black,thick] (1,1) -- (2,0);
\draw[black,thick] (2,2) -- (1,1);
\draw[black,thick] (2,2) -- (4,0);
\filldraw[color=black,fill=green]   (1,1)  circle (\R) node[left,xshift=-0.2cm]{$\scriptstyle #1$};
\filldraw[color=black,fill=green]   (2,2)  circle (\R) node[left,xshift=-0.2cm]{$\scriptstyle #2$};
\end{tikzpicture}}

\newcommand{\orderBA}[2]{\begin{tikzpicture}[scale=0.4]
\tikzset{%
  >=latex, 
  inner sep=0pt,%
  outer sep=2pt,%
  mark coordinate/.style={inner sep=0pt,outer sep=0pt,minimum size=3pt}%
}
\def\R{0.15}
\draw[black,thick] (3,1) -- (2,0);
\draw[black,thick] (3,1) -- (4,0);
\draw[black,thick] (2,2) -- (3,1);
\draw[black,thick] (2,2) -- (0,0);
\filldraw[color=black,fill=green]   (3,1)  circle (\R) node[right,xshift=0.1cm, yshift=0.1cm]{$\scriptstyle #1$};
\filldraw[color=black,fill=green]   (2,2)  circle (\R) node[left,xshift=-0.2cm]{$\scriptstyle #2$};
\end{tikzpicture}}
\newcommand{\orderC}[3]{\begin{tikzpicture}[scale=0.4]
\tikzset{%
  >=latex, 
  inner sep=0pt,%
  outer sep=2pt,%
  mark coordinate/.style={inner sep=0pt,outer sep=0pt,minimum size=3pt}%
}
\def\R{0.15}
\draw[black,thick] (1,1) -- (0,0);
\draw[black,thick] (1,1) -- (2,0);
\draw[black,thick] (2,2) -- (1,1);
\draw[black,thick] (2,2) -- (4,0);
\draw[black,thick] (3,3) -- (2,2);
\draw[black,thick] (3,3) -- (6,0);
\filldraw[color=black,fill=green]   (1,1)  circle (\R) node[left,xshift=-0.2cm]{$\scriptstyle #1$};
\filldraw[color=black,fill=green]   (2,2)  circle (\R) node[left,xshift=-0.2cm]{$\scriptstyle #2$};
\filldraw[color=black,fill=green]   (3,3)  circle (\R) node[left,xshift=-0.2cm]{$\scriptstyle #3$};
\end{tikzpicture}}

\newcommand{\gers}[1]{{\llbracket #1\rrbracket}}

\begin{document}
\pagenumbering{gobble}
\hfill
\vspace{-1.5cm}
\begin{flushright}
    {}
\end{flushright}
\vskip 0.01\textheight
\begin{center}
{\Large\bfseries 
$A_\infty$ Algebras from \\
\vspace{0.4cm}
Slightly Broken Higher Spin Symmetries}

\vskip 0.03\textheight

Alexey \textsc{Sharapov}${}^{1}$ and Evgeny \textsc{Skvortsov}${}^{2,3}$

\vskip 0.03\textheight

{\em ${}^{1}$Physics Faculty, Tomsk State University, \\Lenin ave. 36, Tomsk 634050, Russia}\\
\vspace*{5pt}
{\em ${}^{2}$ Albert Einstein Institute, \\
Am M\"{u}hlenberg 1, D-14476, Potsdam-Golm, Germany}\\
\vspace*{5pt}
{\em ${}^{3}$ Lebedev Institute of Physics, \\
Leninsky ave. 53, 119991 Moscow, Russia}

\end{center}

\vskip 0.02\textheight

\begin{abstract}
We define a class of $A_\infty$-algebras that are obtained by deformations of higher spin symmetries. While higher spin symmetries of a free CFT form an associative algebra, the slightly broken higher spin symmetries give rise to a minimal $A_\infty$-algebra extending the associative one. These $A_\infty$-algebras are related to non-commutative deformation quantization much as the unbroken higher spin symmetries result from the conventional deformation quantization. 
In the case of three dimensions there is an additional parameter that the $A_\infty$-structure depends on,  which is to be related to the Chern--Simons level. The deformations corresponding to the bosonic and fermionic matter lead to the same $A_\infty$-algebra, thus manifesting the three-dimensional bosonization conjecture. In all other cases we consider, the $A_\infty$-deformation is determined by a generalized free field in one dimension lower.

\end{abstract}
\newpage

\tableofcontents
\newpage
\section{Introduction}
\pagenumbering{arabic}
\setcounter{page}{2}
Strong homotopy algebras (SHA), which are also dubbed $A_\infty$ and $L_\infty$ for the cases generalizing associative and Lie algebras, are general enough structures that abstract and formalize what it means to be algebraically consistent in a broad sense. No wonder that many of physical problems can be cast into the framework  of SHA, like string field theory \cite{Witten:1985cc,Zwiebach:1992ie,Gaberdiel:1997ia,Kajiura:2003ax,Erler:2013xta} and the BV-BRST theory of gauge systems \cite{Lada:1992wc,Alexandrov:1995kv,Barnich:2004cr,Hohm:2017pnh}. Even some problems that are seemingly  unrelated to any SHA admit natural solutions by translating them to the SHA setup, e.g. the deformation quantization of Poisson manifolds \cite{Kontsevich:1997vb}. In the present paper, we use the language of SHA in order to describe the slightly broken higher spin symmetries \cite{Maldacena:2012sf,Giombi:2016hkj,Skvortsov:2015pea,Giombi:2016zwa,Aharony:2018npf,Alday:2016jfr,Charan:2017jyc,Yacoby:2018yvy,Sleight:2018ryu} that govern certain nontrivial conformal field theories (CFT) at least in the large-$N$ limit.

It is well known that free CFT's have vast symmetries --- {\it higher spin symmetries} --- that extend the conformal Lie algebra to infinite-dimensional associative algebras, called {\it higher spin algebras} in this context. Higher spin symmetries are related to higher spin currents, i.e., conserved tensors that are bi-linear in the free fields, e.g. $J_s=\phi \pl^s\phi+\ldots$ for the free scalar field $\phi$, with the stress-tensor being just the $s=2$ member of the multiplet. On the other hand, higher spin symmetries generated by abstract conserved tensors $J_s$ are powerful enough as to fix all the correlation functions \cite{Maldacena:2011jn,Boulanger:2013zza,Alba:2013yda,Alba:2015upa} and these turn out to be necessarily given by a free CFT. Therefore, unbroken higher spin symmetries are in one-to-one correspondence with free CFT's. Still it is interesting that the correlators of $J$'s can be directly computed as invariants of higher spin symmetries \cite{Colombo:2012jx,Didenko:2012tv,Didenko:2013bj,Sleight:2016dba, Bonezzi:2017vha}.  

The structure of interacting CFT's is much more complicated. If a given CFT admits a weakly-coupled limit, which is not necessarily free in terms of the fundamental constituents, then one can think of such a CFT as enjoying a {\it slightly broken higher spin symmetry}, the term coined in \cite{Maldacena:2012sf}. In particular, the higher spin currents $J_s$ are not conserved anymore, but their conservation is broken in a very specific way. The examples of main interest include the critical vector model, the Gross--Neveu model and, more generally, the Chern--Simons matter theories in the large-$N$ limit. The last class of models has recently been conjectured to exhibit a number of interesting dualities \cite{Giombi:2011kc, Maldacena:2012sf, Aharony:2012nh,Aharony:2015mjs,Karch:2016sxi,Seiberg:2016gmd}, in particular the three-dimensional bosonization duality. Our expectation is that the dualities can be explained by the slightly broken higher spin symmetries \cite{Maldacena:2012sf} and that this symmetry makes the models exactly soluble, at least in the large-$N$ limit. The purpose of the paper is (i) to define what a slightly broken higher spin symmetry means in mathematical terms since it is not strictly speaking a symmetry in the conventional sense; (ii) to provide an explicit construction; (iii) to explore the simplest consequences including applications to the bosonization duality.

To begin with we would like to stress that higher spin algebras are typically rigid, that is, admit no deformations. Indeed, in $d>2$ the free CFT's are isolated points and do not form continuous families.\footnote{There is a one-parameter family of algebras in $4d$ \cite{Fernando:2009fq,Boulanger:2011se, Manvelyan:2013oua, Joung:2014qya}, the free parameter $\lambda$ being helicity of a free $4d$ conformal field. For non-(half)integer values of the parameter these algebras do not have any natural spacetime and CFT interpretation and for $|\lambda|>1$ there is no local stress-tensor as well. Therefore, only $|\lambda|=0,\tfrac12,1$ correspond to free CFT's. Also, there is a one-parameter family of algebras relevant for higher spin theories in $AdS_3$ whose dual CFT's are $W$ minimal models \cite{Gaberdiel:2010pz}. Lastly, $\mathcal{N}=4$ SYM forms a continuous family of CFT's approaching the free limit at $g=0$. However, for $g\neq0$ the higher spin currents are not conserved and, therefore, SYM does not enjoy higher spin symmetry for $g\neq0$. } The opposite conclusion is also true: higher spin symmetries are the symmetries of free CFT's in $d>2$ \cite{Maldacena:2011jn,Boulanger:2013zza,Alba:2013yda,Alba:2015upa}. Therefore, a slightly broken higher spin symmetry is not about deformation of higher spin algebras as associative (or Lie) algebras. Due to the specific way that higher spin currents $J_s$ fail to be conserved, our proposal is that such deformations fall into the class of $A_\infty$-algebras we construct. 

The $A_\infty$-algebras that describe slightly broken higher spin symmetries are still related to associative algebras and deformations thereof. The class of relevant $A_\infty$-algebras may be of some interest by itself, being closely related to the so-called {\it noncommutative deformation quantization} \cite{Pinczon-1,HOT}. In a few words, suppose that we have an associative (in general non-commutative) algebra and, furthermore, that the product can be deformed as
\begin{align}\label{ast}
    a\ast b=a b +\phi(a,b)\hbar+\cdots\,,\tag{1}
\end{align}
$\hbar$ being a formal deformation parameter. Therefore, we have a one-parameter family of algebras $\mathrm{A}_\hbar$. Here $\phi$ is a Hochschild $2$-cocycle. Construction and classification of such deformations in the case of algebras of smooth functions on Poisson manifolds is the standard problems of deformation quantization that have been solved by Kontsevich \cite{Kontsevich:1997vb} using a string-inspired construction. The deformation problem we are lead to is to promote the formal parameter $\hbar$ to an element of the algebra itself. This clearly has no sense in the realm of usual associative algebras. The idea is to go to the category of $A_\infty$-algebras where it is legitimate to replace $\phi(a,b)\hbar$ with a tri-linear map $m_3(a,b;u)$, with the `deformation parameter' $u$ being now an element of the same algebra. The correspondence principle requires $m_3(a,b;\hbar)=\phi(a,b)\hbar$. This is only the starting point and all the higher structure maps $m_n(a,b;u,\ldots ,w)$ are to be constructed. Taken together, the $m$'s obey the Maurer--Cartan equation and this amounts to defining an $A_\infty$-algebra. One of our results is that the $A_\infty$-structure maps $m_n$ can all be expressed through $\phi$ and the other coefficients in expansion (\ref{ast}). To summarize, given a one-parameter family $\mathrm{A}_\hbar$ of associative algebras  we can explicitly construct an $A_\infty$-algebra that can be viewed as a non-commutative deformation quantization of $\mathrm{A}_\hbar$ at $\hbar=0$:
$$
\begin{tikzcd}
     \text{\parbox{5.0cm}{\centering one-parameter family of \\ associative algebras $\mathrm{A}_\hbar$}}  \arrow[r,"\text{Section 3}"] & [3em] \text{\parbox{5.0cm}{strong homotopy algebra\\ \centering extending $\mathrm{A}_0$}}
\end{tikzcd}
$$
The last step is to establish a relation between the physical realization of the slightly broken higher spin symmetry in vector models and the abstract construction above. When the higher spin symmetry is slightly broken by interactions, the higher spin currents $J_s$ are no longer conserved. Nevertheless, the non-conservation of $J_s$ has a very specific form of double trace operators built out of higher spin currents themselves
\begin{align}\label{intro}
    \pl \cdot J&= g \, [JJ] \tag{2}\,,
\end{align}
where $g$ is a small parameter of order $1/N$. In a sense, higher spin currents are responsible for their own non-conservation.  A remarkable fact is that the multiplet of higher spin currents is isomorphic to the higher spin algebra itself up to a $\mathbb{Z}_2$-twist generated by the inversion map (Section \ref{sec:HScurrents}). Having in mind this correspondence we should be looking for a deformation of the higher spin algebra which is controlled by another element of the algebra (up to the inversion map). Thus, the $\mathbb{Z}_2$-extension of a given higher spin algebra by the inversion map is a useful object to incorporate both the algebra and the higher spin currents. We prove that the $\mathbb{Z}_2$-extended associative algebras are proved to admit at least a one-parameter family of deformations, which we call {\it deformed higher spin algebras}. It is this deformation that is plugged into the general construction of $A_\infty$-algebras just described. The algorithm for constructing the strong homotopy algebra description of the slightly broken higher spin symmetry starting from any higher spin algebra $hs$ looks as follows:
$$
\begin{tikzcd}
    \text{\parbox{1.3cm}{\centering  $hs$ \\ (rigid) }} \arrow[r, "\text{extension}"] &[2em] \text{\parbox{1.5cm}{\centering  $hs \rtimes \mathbb{Z}_2$ \\ (soft) }}  \arrow[r, "\text{deformation}"] & [2em] \text{\parbox{3.0cm}{\centering deformed higher\\  spin algebra}}  \arrow[r] & \text{strong homotopy algebra}
\end{tikzcd}
$$
In words, $hs$ is rigid, but its $\mathbb{Z}_2$-extension, which is needed to incorporate the higher spin currents, is soft and can be deformed into an at least one-parameter family of associative algebras. Lastly, one can use our construction of the strong homotopy algebras out of a one-parameter family of associative algebras (non-commutative deformation quantization).

We also study particular examples of this construction. As is mentioned, higher spin algebras contain the conformal algebra $so(d,2)$ as a Lie subalgebra and have the full information about the spectrum of higher spin currents including the correlation functions. Therefore, they crucially depend on dimension $d$ and on a type of a free CFT they originate from.  Of special interest is the case of three dimensions. Here the structure of the higher spin symmetry breaking is richer than in higher dimensions. Microscopically, this happens due to the presence of an additional parameter related to the level $k$ of the Chern--Simons matter theories. In $d\neq3$ the vector models have a single parameter $N$.\footnote{Note that the breaking of higher spin symmetry in $\mathcal{N}=4$ SYM is different from the one in vector models: the non-conservation equation does not have the form \eqref{intro}. It is the special form of \eqref{intro} that makes higher spin currents close onto themselves at least in the large-$N$ limit. While it would also be interesting to study the breaking of higher spin symmetries in models of matrix type, like SYM and ABJ, we restrict ourselves to models of vector type and use the term {\it slightly broken higher spin symmetry} for vector models only, as introduced in \cite{Maldacena:2012sf}. } The structure of correlation functions is also more complicated with certain parity-odd structures contributing to it \cite{Maldacena:2012sf,Giombi:2011rz,Giombi:2016zwa}. More importantly, the Chern--Simons matter theories with bosonic and fermionic matter seem to describe the same physics and this has lead to the conjecture of the three-dimensional bosonization and related ones \cite{Giombi:2011kc, Maldacena:2012sf, Aharony:2012nh,Aharony:2015mjs,Karch:2016sxi,Seiberg:2016gmd}.

Concerning the three-dimensional bosonization duality, the first observation is that the higher spin algebras of $3d$ free boson and $3d$ free fermion CFT's are isomorphic, which is not the case when $d>3$. This implies that they have to lead to the same $A_\infty$-algebra. Secondly, the deformation that leads to $A_\infty$-algebra is characterized by the second Hochschild cohomology and it turns out to be two-dimensional in $d=3$, while it is one-dimensional in $d>3$. The additional parameter is to be associated with the t'Hooft coupling $\lambda=N/k$. This provides a good evidence for the conjecture to the leading order in $1/N$. The correlation functions should be given by the invariants we discuss at the very end. The appearance of the additional parameter is a feature of the $A_\infty$-algebra, i.e., of the slightly broken higher spin symmetry, and is not seen in the free limit governed by the higher spin algebra. 

To summarize, the unbroken higher spin symmetry is powerful enough\footnote{The restriction to $d>2$ is important, see e.g. \cite{Maldacena:2011jn}, since this result does not hold in $d=2$ and higher spin algebras do not seem to work in $2d$ CFT's the way they do in $d>2$. } in $d\geq3$ as to fix all correlation functions. This property is expected to extend to the more interesting case of the slightly broken higher spin symmetry that underlies a number of nontrivial CFT's at least in the large-$N$ limit. While the former is governed by associative higher spin algebras, the latter leads to the $A_\infty$-algebras we propose in the paper. These $A_\infty$-algebras still are fully controlled by {\it deformed higher spin algebras}. The invariants of these algebras should give the correlation functions, much as they do for the usual higher spin algebras. The relation between some of the structures on the physics and mathematical sides is illustrated by the following diagrams:
$$
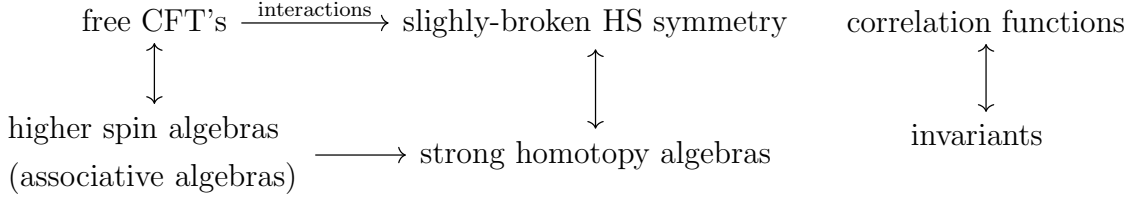
\begin{tikzcd}
    \text{free CFT's} \arrow[d,leftrightarrow] \arrow[r, "\text{interactions}"] & \text{slighly-broken HS symmetry}  \arrow[d,leftrightarrow] \\
    \text{\parbox{3.9cm}{higher spin algebras\\ (associative algebras)}} \arrow[r] & \text{strong homotopy algebras}
\end{tikzcd}\quad
\begin{tikzcd}
     \text{correlation functions}\arrow[d,leftrightarrow]\\ [5pt]
     \parbox{2cm}{invariants\\ \phantom{blah}}
\end{tikzcd}
$$ 

The rest of the paper is organized as follows. We begin in Section \ref{sec:Adefinition} with the definition of $A_\infty$-algebras in terms of the Gerstenhaber bracket. In Section \ref{sec:Aconstruction}, we define and construct a class of $A_\infty$-algebras that can be thought of as non-commutative deformation quantization of associative algebras. Various definitions and examples of higher spin algebras are recalled in Section \ref{sec:HSA}. In Section \ref{sec:SlightlyBroken} we define the $A_\infty$-algebra of the slightly broken higher spin symmetry. Some explicit oscillator realizations of these deformations are discussed in Section \ref{sec:Oscillators}. Conclusions are in Section \ref{sec:Conclusions}. Several appendices are devoted to more technical aspects, in particular in Appendix \ref{sec:HSAdeformed} we prove that certain simple extensions of higher spin algebras admit deformations.

\section{\texorpdfstring{$\boldsymbol{A_\infty}$}{A-infinity} Algebras}
\label{sec:Adefinition}
There are several equivalent definitions of $A_\infty$-algebras: (i) via Stasheff's relations \cite{Stasheff}; (ii) via a nilpotent coderivation on the tensor coalgebra of the suspended graded algebra and (iii) via the Gerstenhaber bracket. Throughout the paper we will exclusively use the last one. 

Let $V$ be a $\mathbb{Z}$-graded vector space~$V=\bigoplus_k V^k$. Consider the space $Hom(TV,V)$ of all maps from the tensor algebra $TV=\bigoplus_n T^nV$ of $V$ to the space $V$ itself. The element of $Hom(T^nV,V)$, called {\it $n$-cochains}, are multilinear functions $f(a_1,a_2,\ldots,a_n)$ on $V$. The $\mathbb{Z}$-grading on $V$ induces that on $Hom(TV,V)$; by definition, 
$$
|f|=|f(a_1,a_2,\ldots,a_n)|-\sum_{k=1}^n|a_k|\,.
$$
The  $\circ$-product of an $n$-cochain $f$ and an $m$-cochain $g$ is a natural operation that nests one map into the other with the usual Koszul signs
\begin{align}\label{comp-prod}
    \begin{aligned}
    (f\circ g)&(a_1\otimes a_2\otimes\cdots\otimes a_{m+n-1} )=\\
     &=\sum_{i=0}^{n-1}(-1)^{|g|\sum_{j=1}^i|a_j|} f(a_1\otimes \cdots\otimes a_i\otimes g(a_{i+1}\otimes \cdots\otimes a_{i+m})\otimes \cdots \otimes a_{m+n-1})\,.
    \end{aligned}
\end{align}
It should be noted that the $\circ$-product is non-associative. Nevertheless, the following bracket, called {\it Gerstenhaber bracket},
\begin{align}
   \gers{f,g}=f\circ g-(-1)^{|f||g|}g\circ f\,,
\end{align}
is graded skew-symmetric and obeys the Jacobi identity: 
\begin{align}
   \gers{f,g}&=-(-1)^{|f||g|}\gers{g,f} \,, &&\gers{\gers{f,g},h}=\gers{f,\gers{g,h}}-(-1)^{|f||g|}\gers{g,\gers{f,h}}\,.
\end{align}
Given a $\mathbb{Z}$-graded space $V$ and a sum $m=m_1+m_2+\cdots $ of degree-one maps $m_n: T^nV\rightarrow V$, the $A_\infty$-structure is defined simply as a solution to the Maurer--Cartan equation:
\begin{align}
    \gers{m,m}&=0\,.
\end{align}
Upon expansion  $m=m_1+m_2+\cdots $ the first few relations have a simple interpretation: $m_1$ is a differential, $m_1m_1=0$; $m_2$ is a bi-linear product differentiated by $m_1$ by the graded Leibniz rule  
\begin{align}
-m_1 m_2(a,b)=m_2(m_1(a),b)+(-)^{|a|}m_2(a,m_1(b))\,.    
\end{align}
However, $m_2$ is not associative in general, associativity is true up to a coboundary controlled by $m_3$:
\begin{align*}
m_2(m_2(a,b),c)+&(-)^{|a|}m_2(a,m_2(b,c))+m_1 m_3(a,b,c)+m_3(m_1(a),b,c)+\\
&\qquad+(-)^{|a|}m_3(a,m_1(b),c)+(-)^{|a|+|b|}m_3(a,b,m_1(c))=0\,.
\end{align*}
{\it NB: it is common in the literature to define $A_\infty$-algebras via maps on the suspension $V[1]$ of the corresponding graded space $V$. Then $m_n$ has degree $2-n$. We prefer to prepare the 'experimental setup' in such a way that $V$ is already suspended. This prevents appearance of many sign factors and all $m_n$ have now degree one. For example, an associative algebra $A$ is understood as a graded algebra with the only nonzero component leaving in degree $-1$, so that  multiplication is a degree-one map $m_2$ taking $A_{-1}\otimes A_{-1}$ to $A_{-1}$. As a consequence of such a degree assignment the associativity condition has the right form} $$ \gers{m_2,m_2}(a,b,c)=2m_2(m_2(a,b),c)-2m_2(a,m_2(b,c))=0\,.$$
Certain $A_\infty$-algebras deserve their own names. {\it Minimal} $A_\infty$-algebras do not have the lowest map $m_1$, i.e., differential. Such algebras arise naturally when passing to the cohomology $H(m_1)$ of $m_1$ and dragging the $A_\infty$-structures there, the resulting algebras are called {\it minimal models}, see e.g. \cite{Kajiura:2003ax}. {\it Differential graded algebras} (DGA) have only $m_1$ and $m_2$, i.e., a differential and a bi-linear product that respect the Leibniz rule.

Note that for a genuine $A_\infty$-structure to arise it is necessary that $V$ has more than one graded component due to the degree requirement. The only possibility with just one nontrivial component is $V=V_{-1}$, then $m_2$ is just an associative product on $V$.

\section{\texorpdfstring{$\boldsymbol{A_\infty}$}{A-infinity} from Deformations of Associative Algebras}
\label{sec:Aconstruction}
In this section, we construct an $A_\infty$-algebra out of a one-parameter family of associative algebras. Even though the construction is inspired by the study of the slightly broken higher spin symmetry, it is quite general and may be of independent interest as a new way to build a large class of $A_\infty$-algebras. There are certain special properties of HSA that allows one to describe the corresponding $A_\infty$-algebras in more detail and there are tools to explicitly construct them, which will be discussed in Sections \ref{sec:Oscillators} and Appendix \ref{sec:HSAdeformed}. Throughout this section, we let $\ass$ denote any associative algebra.

Given an associative algebra $\ass$, it is clear that due to the restrictions imposed by the grading, there cannot be any interesting $A_\infty$-structure on it; the only possibility is to deform $\ass$ itself as an associative algebra. We define the $A_\infty$-structure perturbatively and the first step is to extend $\ass$ by any its bimodule $M$; in so doing, $\ass$ and $M$ are prescribed the degrees $-1$ and $0$, respectively. At the lowest order the $A_\infty$-structure is simply equivalent to the definitions above: there is only $m_2$ that is defined for various pairs $A_{-1}\otimes A_{-1}$ (the $\ass$ product), $A_{-1}\otimes A_0$ (the left action of $\ass$ on $M$), $A_{0}\otimes A_{-1}$ (the right action of $\ass$ on $M$). All these conditions are summarize by the Stasheff identity:
\begin{align}
    m_2(m_2(a,b),c)+(-)^{|a|}m_2(a,m_2(b,c))&=0 &&\Longleftrightarrow && \gers{m_2,m_2}=0\,.
\end{align}
Denoting elements of $A_{-1}$ by $a,b,\ldots$ and elements of $A_0$ by $u,v,\ldots$ we have \footnote{The left/right action is denoted by multiplication, $au$ and $ua$. }
\begin{align}\label{m2form}
    m_2(a,b)&= ab\,, &m_2(a,u)&=au\,, & m_2(u,a)&=-ua\,, & m_2(u,v)&=0\,.
\end{align}
Now one tries to deform this rather trivial $A_\infty$-structure and the first-order deformations $m^{(1)}$ can be described in terms of the Hochschild cohomology of $\ass$. Introducing  the Hochschild differential $\delta=\gers{m_2,\bullet}$, one can identify the nontrivial first-order deformations $m^{(1)}$ with the nontrivial $\delta$-cocycles, 
\begin{align}
    \delta m^{(1)}&=0 && \Longleftrightarrow&& \gers{m_2, m^{(1)}}=0\,.
\end{align}
In other words, the space of infinitesimal deformations is identified with the $\delta$-cohomology in degree $1$, while the second $\delta$-cohomology group is responsible for possible obstructions to deformation. 

The first-order deformation should have the form $m^{(1)}=m_3(\bullet,\bullet,\bullet)$ with arguments from $A_{-1}$ and $A_0$. Various homogeneous  components of $\delta m_3=0$ are collected in Appendix \ref{app:FirstOrder}, while the first and the last ones are:
\begin{align}
    -am_3(b,c,u)+m_3(ab,c,u)-m_3(a,bc,u)+m_3(a,b,cu)&=0\,,\\
    \ldots&=0\,,\\
    m_3(u,a,b)v+um_3(a,b,v)+m_3(ua,b,v)+m_3(u,ab,v)-m_3(u,a,bv)&=0\,.
\end{align}
For any associative algebra $\ass$ there is at least one natural bimodule, that is, $\ass$ itself. Let us take $A_0$ to be $\ass$, in which case the deformation can be described in more detail. If $\ass$ admits a deformation as an associative algebra, then the second Hochschild cohomology group is nonzero, $HH^2(\ass,\ass)\neq 0$. Given an element $[\phi]\in HH^2(\ass,\ass)$ represented by a cocycle $\phi$, the standard deformation of the associative structure reads
\begin{align}
    a\ast b&= ab+ \phi(a,b) \hbar+\mathcal{O}(\hbar^2)\,,
\end{align}
where the deformation parameter $\hbar$ can live in the base field or even in the center of $\ass$. If the deformation is unobstructed, we can construct a one-parameter family of algebras $\ass_\hbar$ that starts at $\ass$ for $\hbar=0$. When $A_0\sim \ass$ the $A_\infty$-algebra we are trying to construct upgrades the deformation parameter $\hbar$ to an element of $A_0$. The observation is that for $A_0\sim \ass$ one can always put  
\besubeqs\label{formA}
\begin{align}
    m_3(a,b,u)&=\phi(a,b)u\,, && m_3(a,u,v)=\phi(a,u)v\,,\\
    m_3(a,u,b)&=0\,, && m_3(u,a,v)=-\phi(u,a)v \,,\\
    m_3(u,a,b)&=0\,, && m_3(u,v,a)=0\,.
\end{align}
\esubeqs
Here the `deformation parameter' $u\in \ass$ was placed on the right in $m_3(a,b,u)$. It is also possible to place it on the left
\besubeqs\label{formB}
\begin{align}
    m_3(a,b,u)&=0\,, && m_3(a,u,v)=0\,,\\
    m_3(a,u,b)&=0\,, && m_3(u,a,v)=u\phi(a,v)\,,\\
    m_3(u,a,b)&=u\phi(a,b)\,, && m_3(u,v,a)=-u\phi(v,a)\,.
\end{align}
\esubeqs
For $u$ in the base field (or more generally in the center of $\ass$) the left $u\phi(a,b)$ and the right $\phi(a,b)u$ deformations are clearly equivalent. This property extends to the $A_\infty$-structure, namely, the left and right deformations differ from  each other by a  trivial deformation $m_3=\delta g$, where
$g(a,u)=\phi(a,u)$.

The $A_\infty$-algebra we are  constructing extends the deformation parameter $\hbar$ to an element of $A_0$, which may be the algebra itself (or its bimodule). This is usually referred to as deformation with noncommutative base. If such an $A_\infty$-algebra can be constructed, it admits a truncation where $A_0$ is replaced by the center ${Z}(\ass)$, or just by $\hbar$, that is closely related to the one-parameter family of algebras $\ass_\hbar$. 

\subsection{Explicit Construction}
The central statement of the present paper is that the $A_\infty$-structure of the previous section, is fully determined by the deformation of the underlying associative algebra. Assuming that the deformed product 
\begin{align}\label{absdeformed}
    a\ast b&= ab+\sum_{k>0}\phi_k(a,b)\hbar^k
\end{align}
is known, we give an explicit formula for all $m_n$. The defining relation for the $A_\infty$-structure, i.e., the Maurer--Cartan equation 
\begin{align}
   \gers{m,m}&=0 && \Longleftrightarrow && \delta m_n +\sum_{i+j=n+2}m_i \circ m_j=0\,,
\end{align}
is satisfied as a consequence of the associativity of the deformed product
\begin{align}
   a\ast (b\ast c)-(a\ast b)\ast c&=0 && \Longleftrightarrow && \delta \phi_n+\sum_{i+j=n-1}\phi_i\circ\phi_j=0\,.
\end{align}
Here $\delta=\gers{m_2,\bullet}$ is the Hochschild differential associated to the undeformed product \eqref{m2form}. We have three equivalent forms of $m_n$: recursive, in terms of binary trees, and through generating equations. Let us discuss them in order. 

In general, there are two types of ambiguities in the definition of $m_n$. (i) As usual in  deformation quantization, one can redefine the deformed product $\ast$ via a linear change of variables $a\rightarrow D(a)=a+\sum_k D_k(a)\hbar^k$. Then, the new product is given by $D(D^{-1}(a)\ast D^{-1}(b))$. (ii) One can perform various redefinitions at the level of $A_\infty$-structure, which is done by exponentiating the infinitesimal gauge transformation
\begin{align}
    \dot m(t)&=\gers{m(t),\xi}\,, && m(0)=m\,,
\end{align}
for some cochain $\xi$ of degree zero. 
The $A_\infty$ gauge transformations are more general than redefinitions of the associative product. We have observed that the $A_\infty$-transformations allow one to cast the first-order deformation into the right form (with all, or all but one, $A_{0}$-factors staying on the right):
\begin{align}
    m_3(a,b,u)=f_3(a,b)u \,,&&m_3(a,u,v)=f_3(a,u)v\,, && m_3(u,a,v)=-f_3(u,a)v\,,
\end{align}
and all other orderings of $a,b, u,v$ in $m_3$ give zero result. Here $f_3(a,b)=\phi_1(a,b)$ is determined by the first-order deformation in \eqref{absdeformed}. The full solution can be sought for in a similar form:
\begin{align}
    m_n(a,b,u,\ldots,v)&=+f_n(a,b,u,\ldots)v\,,\\
    m_n(a,u,\ldots,v,w)&=+f_n(a,u,\ldots,v)w\,,\\
    m_n(u,a,\ldots,v,w)&=-f_n(u,a,\ldots,v)w\,.
\end{align}
Therefore, the problem is reduced to defining one function $f_n$ of $(n-1)$ arguments per each set of structure maps $m_n$ with only three orderings being nontrivial. It is not hard to see that the equation for $m_4$, $\delta m_4 +m_3\circ m_3=0$,  
is solved by
\begin{align}\label{ffour}
    f_4(a,b,u)&=\phi_2(a,b)u +\phi_1(\phi_1(a,b),u)\,.
\end{align}
At the next order we have to solve $\delta m_5 +m_3\circ m_4+ m_4\circ m_3=0$, which is  satisfied by 
\begin{align}
\begin{aligned}
    f_5(a,b,u,v)&=\phi_1(\phi_1(\phi_1(a,b),u),v)+\phi_2(\phi_1(a,b),u)v+\phi_1(\phi_2(a,b),u)v+\\
    &\qquad \qquad+\phi_1(\phi_2(a,b)u,v)+\phi_3(a,b)uv\,.
\end{aligned}
\end{align}
The following graphical representation can be useful. We consider planar binary trees with vertices labelled by $0,1,2,\ldots$. A vertex with label $k$ corresponds to $\phi_k$ and the two incoming edges correspond to the arguments. Functions $f_3$, $f_4$ and $f_5$ can then be depicted as
\begin{align*}
    f_3&=\orderA{1}\,,\\
    f_4&=\orderB{1}{1}+\orderB{2}{0}\,,\\
    f_5&=\orderC{1}{1}{1}+\orderC{1}{2}{0}+\orderC{2}{1}{0}+\orderC{2}{0}{1}+\orderC{3}{0}{0}\,.
\end{align*}
\paragraph{Solution, recursive formula.} In order to write down a recursive formula for $f_n$ let us introduce some further notation. It is clear that any $f_n$ can be decomposed according to the number of the multiplicative arguments on the right:
\begin{align*}
    f_n(a,b,u,\ldots,v,w)&=f_{n,0}(a,b,u,\ldots,v,w)+f_{n,1}(a,b,u,\ldots,v)w+f_{n,2}(a,b,u,\ldots)vw+\ldots\,.
\end{align*}
There is an associated filtration, where the leftover $r_{n,k}$ contains all the terms in the decomposition with at least $k$ multiplicative arguments  on the right:
\besubeqs
\begin{align}
    f_n(a,b,\ldots,v,w)&\equiv r_{n,0}(a,b,\ldots,v,w)\\
        &=f_{n,0}(a,b,\ldots,v,w)+r_{n,1}(a,b,\ldots,v)w\\
        &=f_{n,0}(a,b,\ldots,v,w)+f_{n,1}(a,b,\ldots,v)w+r_{n,2}(a,b,\ldots)vw\,, \quad \text{etc.}
\end{align}
\esubeqs
Our claim is that all $f_n$ are obtained by means of the following recursive relations:\footnote{It is useful to define $f_2= r_{2,0}$ as the identity map.}
\besubeqs
\begin{align}
     f_{n,0}&=\phi_1(r_{n-1,0},\bullet)\,,\\
    f_{n,1}&=\phi_2(r_{n-2,0},\bullet)+\phi_1(r_{n-1,1},\bullet)\,,\\
    f_{n,2}&=\phi_3(r_{n-2,0},\bullet)+\phi_2(r_{n-2,1},\bullet)+\phi_1(r_{n-3,0},\bullet)\,, \\
    \cdots &\\
    f_{n,k}&= \sum_{i=0}^{i=k} \phi_{k-i+1}(r_{n-k+i-1,i},\bullet)\,.
\end{align}
\esubeqs
The formulae above together with the initial condition $f_3=\phi_1$ allow one to  reconstruct the $A_\infty$-structure, $m_n$, in terms of the bi-linear maps $\phi_{k}$ defining the $\ast$-product (including the initial product $\phi_0(a,b)=ab$). 

While $f_n$'s are, in general, quite complicated functions with nested $\phi_k$, there are some general properties that are easy to see. (a) The first and the last terms in $f_n$ are of the form
\begin{align}
    f_n(a,b,u,\ldots,v,w)&=\phi_1(\phi_1(\ldots(\phi_1(a,b),u),\ldots,v),w)+\cdots+\phi_{n-2}(a,b)u\ldots vw\,.
\end{align}
The presence of the last term is obvious as for $u,\ldots,v,w$ in the base field the deformation should reduce to the deformed product\footnote{We should assume here that the deformation of the product is properly normalized, $\phi_k(a,1)=0$.}
\begin{align}
    f_n(a,b,\hbar,\ldots ,\hbar,\hbar)&=\phi_{n-2}(a,b)\hbar^{n-2}\,.
\end{align}
(b) The graphs that show up in the decomposition of $f_n$ are all left-aligned, i.e., are the simplest ones with all edges emerging from just one branch on the left. Such graphs can be parameterized by a sequence of numbers listing the indices of the vertices when read from left to right, e.g. $(2,0)$ and $(1,1)$ for $f_4$. Such a simple form is the consequence of a particular $A_\infty$ gauge we chose. By performing an $A_\infty$ gauge transformations one can arrive at various other forms. In particular, there exists the right-aligned form, which is obtained by reflection of the graphs. (c) All graphs contributing to $f_n$ have the total weight $n-2$, where the weight is the sum over the indices of the vertices in a graph. 
(d) Not all possible left-aligned graphs with a correct weight contribute to the expansion of $f_n$. All admissible graphs enter with multiplicity one.  

\paragraph{Solution, explicit formula.} Instead of the recursive definition given above it is possible to describe the set of trees that contribute to $f_n$ in a more direct way. This is easier to do in terms of the sequences of natural numbers 
\begin{align}
    (m_k,l_k,\ldots ,m_1,l_1)
\end{align}
that correspond to the trees encoded by the weights 
\begin{align}
    &m_k+1,\overbrace{0,\ldots,0}^{l_k},m_{k-1}+1,0,\ldots ,0,m_2+1,\overbrace{0,\ldots ,0}^{l_2}m_1+1,\overbrace{0,\ldots ,0}^{l_1} \,,
\end{align}
or, pictorially,
\begin{align}
f_n(a,b,u,\ldots,w)&\ni \quad
\parbox{4cm}{\begin{tikzpicture}[scale=0.4]
    \def\R{0.15}
    \draw[black,thick] node[below]{$\scriptstyle a$} (0,0) -- (3.5,3.5);
    \draw[black,thick] (1,1) -- (2,0) node[below]{$\scriptstyle b$};
    \filldraw[color=black,fill=green]   (1,1)  circle (\R) node[left,xshift=-0.2cm]{$\scriptstyle m_k+1$};
    \draw[black,thick] (1.5,1.5) -- (2,1.0) node[below right]{$\scriptstyle u$};
    \draw[black,thick] (2,2) -- (2.5,1.5);
    \draw[black,thick] (2.5,2.5) -- (3,2);
    \draw[black,thick] (3,3) -- (4,2);
    \filldraw[color=black,fill=green]   (3,3)  circle (\R); 
    \draw[black,thick] (3.9,3.9) -- (6.5,6.5);
    \draw[black,thick] (4.5,4.5) -- (5,4);    
    \draw[black,thick] (5,5) -- (6,4);
    \filldraw[color=black,fill=green]   (5,5)  circle (\R) node[left,xshift=-0.2cm]{$\scriptstyle m_{1}+1$};    
    \draw[black,thick] (5.5,5.5) -- (6,5);
    \draw[black,thick] (6,6) -- (6.5,5.5);    
    \draw[black,thick] (6.5,6.5) -- (7,6) node[below right]{$\scriptstyle w$};
    \draw[snake=brace,raise snake=5pt,black, thick] (5.5,5.5) -- (6.5,6.5) node[left, xshift=-0.25cm, yshift=0.15cm]{$\scriptstyle l_1$};
    \draw[snake=brace,raise snake=5pt,black, thick] (1.5,1.5) -- (2.5,2.5) node[left, xshift=-0.25cm, yshift=0.15cm]{$\scriptstyle l_k$};   
\end{tikzpicture}}
\end{align}
Here the edges corresponding to the multiplicative arguments on the right are drawn a bit shorter. Some of the arguments are displayed.

Equivalently, every such sequence corresponds to the expression
\begin{align}
    \phi_{m_1+1}\left( \ldots \phi_{m_{k-1}+1}(\phi_{m_k+1}(\mathds{1},\mathds{1})\mathds{1}^{l_k},\mathds{1}){\mathds{1}}^{l_{k-1}},\ldots,{\mathds{1}}^{l_2},\mathds{1}\right)\mathds{1}^{l_1}(a_0\otimes\ldots\otimes a_{n-2})\,,
\end{align}
where $\mathds{1}$ is the identity map. In this notation $l_i$ is the number of the multiplicative arguments on the right at level $i$ and $m_i$ stands for the insertion of $\phi_{m_i+1}$. Now we need to specify which of the sequences or trees are admissible. They satisfy
\begin{align*}
    l_1&\in [0,n-2-k]\,, & m_1&\in [0,l_1]\,,\\
    l_2&\in [0,n-2-k-l_1]\,, & m_2&\in [0,l_1+l_2-m_1]\,,\\
        \cdots\\
    l_k&\in [0,n-2-k-l_1-\cdots-l_{k-1}]\,, & m_k&\in [0,l_1+...+l_k-m_1-\cdots-m_{k-1}]\,.
\end{align*}
Equivalently, all the terms (trees) contributing to $f_n$ can be enumerated via pairs of Young diagrams. One should write down all possible Young diagrams with the first row of length $n-2-k$ for all $k=1,\ldots,n-2$ and with $k$ rows.  Given such a diagram, one should write down all possible subdiagrams such that the first row is of the same length $n-2-k$.  Any such pair of Young diagrams gives a sequence of $l_i$ and $m_i$ that are admissible. Some of $l_i$ and $m_i$ can be zero, provided that the Young diagram is a proper one (the length of the rows is nondecreasing upwards). Pictorially, such pairs look as follows:
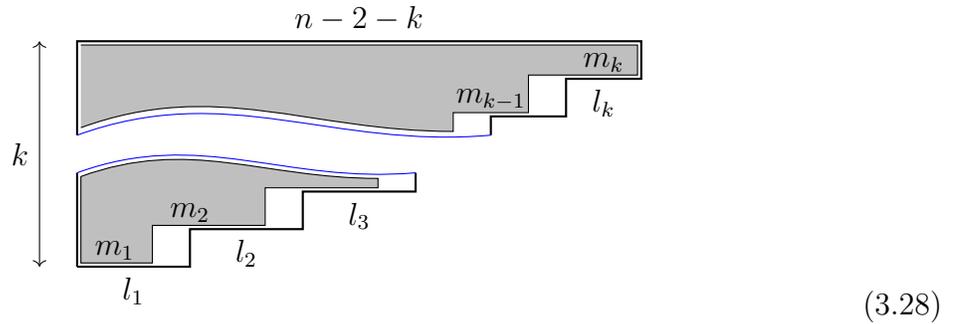
\begin{align}
    \begin{tikzpicture}[scale=0.5]
        \draw[thick, black] (0,0) -- (3,0) -- (3,1) -- (6,1) -- (6,2) -- (9,2) -- (9,2.5);
        \draw[thick, black] (0,0) -- (0,2.5);
        \draw[thick, black] (0,3.5) -- (0,6) -- (15,6) -- (15,5) -- (13,5) -- (13,4) -- (11,4) -- (11,3.5);
        \draw[blue] (0,2.5) to [out=20,in=185] (9,2.5);
        \draw[blue] (0,3.5) to [out=20,in=185] (11,3.5);
        \node[below] at (1.5,0) {$l_1$};
        \node[above] at (7.5,6) {$n-2-k$};
        \node[below] at (4.5,1) {$l_2$};
        \node[below] at (7.5,2) {$l_3$};
        \node[below] at (14,5) {$l_k$};
        \draw[<->] (-1,0) -- (-1,6);
        \node[left] at (-1,3) {$k$};
        \draw[fill=lightgray] (0.1,2.4) to [out=20,in=180] (8,2.35) -- (8,2.1) -- (5,2.1) -- (5,1.1) -- (2, 1.1) -- (2,0.1) -- (0.1, 0.1) -- (0.1,2.4) ;
        \node[above, yshift=-0.1cm] at (1,0.1) {$m_1$};
        \node[above, yshift=-0.1cm] at (3,1.1) {$m_2$};
        \draw[fill=lightgray]  (0.1,3.7) to [out=20,in=180] (10,3.6) -- (10,4.1) -- (12, 4.1) -- (12,5.1) -- (14.9, 5.1) -- (14.9, 5.9) -- (0.1, 5.9) ;
        \node[above, yshift=-0.1cm] at (14,5.1) {$m_k$};
        \node[above, yshift=-0.1cm] at (11,4.1) {$m_{k-1}$};
\end{tikzpicture}
\end{align}
For example, the pair of empty diagrams $(\bullet,\bullet)$ means $k=n-2$, $l_{1,...,k}=m_{1,...,k}=0$ and corresponds to
\begin{align}
    \phi_1(\ldots\phi_1(\phi_1(a_0,a_1),a_2),\ldots,a_{n-2})\,.
\end{align}
The one-row Young diagram of length $n-3$ implies that $k=1$, $m_1=l_1=n-3$ and corresponds to
\begin{align}
    \phi_{n-2}(a_0,a_1)a_2\ldots a_{n-2}\,.
\end{align}
In this language the expansions for $f_4$, $f_5$ and $f_6$ can be written as
\begin{align*}
    f_4&= \left(\bullet,\bullet\right)\oplus \left(\YoungmA,\YoungmA\right)\,,\\
    f_5&= f_4\oplus\left(\YoungmB,\YoungmB\right)\oplus\left(\YoungmAA,\YoungmAA\right)\oplus\left(\YoungmAA,\YoungmA\right)\,,\\
    f_6&= f_5
    \oplus\left(\YoungmC,\YoungmC\right)\oplus\left(\YoungmBA,\YoungmBA\right)\oplus\left(\YoungmBA,\YoungmB\right)\oplus\left(\YoungmBB,\YoungmBB\right)\oplus\left(\YoungmBB,\YoungmBA\right)\oplus\left(\YoungmBB,\YoungmB\right)\oplus\left(\YoungmAAA,\YoungmAAA\right)\oplus\left(\YoungmAAA,\YoungmAA\right)\oplus\left(\YoungmAAA,\YoungmA\right).
\end{align*}

\paragraph{Solution, generating equation.} A combinatorial proof that the two forms above do solve our problem is sketched in Appendix \ref{app:Proof}. Nevertheless, it is desirable to get all $m$'s in a way that makes their existence obvious. To this end, we recall the construction of braces, which were first introduced in \cite{Kadeishvili98} (see also \cite{Getzler93, GV}). A $k$-brace is a multi-linear map that assigns to any set of $k+1$ Hochschild cochains $f, g_1,\ldots g_k$ a new cochain $f\{g_1,\ldots ,g_k\}$ defined by the rule 
\begin{align}
    f\{g_1,\ldots ,g_k\}(a_1,\ldots)&=\sum \pm f(a_1,\ldots,g_1(\ldots),\ldots,g_2(\ldots),\ldots,g_k(\ldots),\ldots)\,.
\end{align}
Here the cochains $g_i$ are inserted as arguments into the cochain $f$ and the sum is over all unshuffles (i.e., the order of $g_i$ is preserved) with natural signs (whenever $g_i$ has to jump over $a_j$ an obvious sign $(-)^{|g_i||a_j|}$ is generated). For $k=1$ we get the Gerstenhaber $\circ$-product (\ref{comp-prod}), that is, $f\{g\}=f\circ g$.

As was shown in \cite{Getzler93}, any $A_\infty$-structure $m$ on $V$ can be lifted to an $A_\infty$-structure $M=M_1+M_2+\cdots$ on the space of Hochschild  cochains $Hom (TV,V)$ by setting 
\begin{equation}
M_1(g_1)=\gers{m,g_1}\,,\qquad  M_k(g_1,\ldots,g_k)=m\{g_1,\ldots, g_k\}\,,\qquad k>1\,.
\end{equation}
Using the properties of the braces, one can find \cite{Getzler93, GV}
\begin{equation}
    \gers{M,M}(g_1,\ldots)=\gers{m,m}\{g_1,\ldots\}=0\,.
\end{equation}
In other words, $M$ satisfies the Maurer--Cartan equation whenever $m$ does so. Expanding the former structure in homogeneous components, $M$, one gets an infinite sequence of relations
$$
\gers{M_1,M_1}=0\,,\qquad \gers{M_1,M_2}=0\,,\qquad \ldots\,.
$$
As is seen  the first term $M_1$ defines a differential $D=\gers{m,\bullet}$ on the space $Hom(TV,V)$. The second relation takes then the form 
\begin{equation}
D M_2(g_1,g_2)+M_2(Dg_1,g_2)+(-1)^{|g_1|} M_2(g_1, Dg_2)=0\,.
\end{equation}
In particular, this means that $M_2$ maps any pair of $D$-cocycles $g_1$ and $g_2$ to a $D$-cocycle $M_2(g_1,g_2)$.  

Suppose now that we are given a two-parameter family $m=m(\hbar,s)$ of $A_\infty$-structures on $V$. Then, differentiating the defining condition $\gers{m,m}=0$ by the parameters, one readily concludes that the partial derivatives $\partial_\hbar m$ and $\partial_s m$ are $D$-cocycles for all $\hbar$ and $s$. Indeed, 
\begin{equation}
    D\partial_\hbar m=\gers{m,\partial_\hbar m}=\frac12\partial_\hbar\gers{m, m}=0\,,\qquad D\partial_s m=\gers{m,\partial_s m}=\frac12 \partial_s\gers{m, m}=0\,. 
\end{equation}
Applying to them $M_2$ yields then one more  family of $D$-cocycles $$M_2(\partial_\hbar m, \partial_s m)=m\{\partial_\hbar m, \partial_s m\}\,.$$ 
We can increase the number  of parameters entering $m$ by considering the flow in the space of cochains 
\begin{align}\label{mEq}
    \pl_t m=m\left\{\pl_\hbar m, \pl_s m\right\}
\end{align}
with respect to the `time' $t$.   Solutions to this equation form a three-parameter family of the cochains $m(t,\hbar,s)$. A simple observation is that the flow (\ref{mEq}) can be consistently restricted to the surface $\gers{m,m}=0$  identified with the set of Maurer--Cartan elements.  Indeed, denoting $L=\gers{m,m}$, we find  
\begin{equation}
\partial_t L=2\gers{m, m\left\{\pl_\hbar m, \pl_s m\right\}}=-\gers{\pl_\hbar L, \pl_s m} +\gers{\pl_\hbar m, \pl_s L}\,.
\end{equation}
Hence, choosing initial data  $m(0,\hbar,s)$ for the solutions to Eq. (\ref{mEq}) on the surface  $L=0$, we will get three-parameter families $m(t,\hbar,s)$ of the Maurer--Cartan elements. Let us take 
\begin{equation}\label{in-d}
  m(0,\hbar,s)= \mu(\hbar)+s\partial\,.
\end{equation}
Here the parameter $s$ is prescribed the degree $2$; $\pl$ is the degree $-1$ differential on $A_{-1}\oplus A_0$ that maps $A_{0}$ to $A_{-1}$ as identity isomorphism and maps $A_{-1}$ to $0$, which is essentially a formal way to retract an element from the bimodule and reinterpret it as an element of the algebra again; $\mu(\hbar)$ is the bimodule structure with respect to the full deformed product \eqref{absdeformed}:  $$\mu(\hbar)(a,b)=a\ast b\,,\qquad \mu(\hbar)(a,u)=a\ast u\,,\qquad \mu(\hbar)(u,a)=-u\ast a\,.$$ 
The Maurer--Cartan equation for (\ref{in-d}) is equivalent to the relations
\begin{equation}
    \gers{\mu(\hbar),\mu(\hbar)}=0\,,\qquad \gers{\mu(\hbar),\partial}=0\,,\qquad \gers{\partial,\partial}=0\,,
\end{equation}
which are obviously satisfied. Notice that both (\ref{in-d}) and the r.h.s. of (\ref{mEq}) are of degree $1$; hence, so is the solution $m(t,\hbar,s)$ to Eq. (\ref{mEq}) with the initial condition (\ref{in-d}).

Now, all $m_n$ can be generated systematically by solving \eqref{mEq} order by order in $t$, $m=m_2+tm_3+t^2 m_4+\ldots$, and setting $\hbar=s=0$ at the end. For example, at the first-order we find
\begin{align}
    m_3&= \mu\{ \mu', \pl \} && \longrightarrow && 
    \begin{cases}
        \mu\{ \mu', \pl \}(a,b,u)=+\mu(\mu'(a,b), \pl(u))\stackrel{\scriptstyle{\hbar=0}}{=}+\phi_1(a,b)u\,,\\
        \mu\{ \mu', \pl \}(a,u,v)=-\mu(\mu'(a,u), \pl(v))\stackrel{\scriptstyle{\hbar=0}}{=}+\phi_1(a,u)v\,,\\
        \mu\{ \mu', \pl \}(u,a,v)=-\mu(\mu'(u,a), \pl(v))\stackrel{\scriptstyle{\hbar=0}}{=}-\phi_1(u,a)v\,,
    \end{cases}
\end{align}
where on the right we evaluated the map on the left for various triplets of arguments. At the second order we obtain the relation
\begin{align}
   2 m_4&= m_3\{ \pl_\hbar \mu, \pl \}+\mu\{\pl_\hbar m_3, \pl\}\,,
\end{align}
and hence
\begin{align}
   m_4(a,b,u,v)&=\mu(\mu'(\mu'(a,b), \pl(u)), \pl(v))+\tfrac12\mu(\mu(\mu''(a,b), \pl(u)),\pl(v))=\\
   &\stackrel{\scriptstyle{\hbar=0}}{=} \phi_1(\phi_1(a,b),u)v+ \phi_2(a,b)uv\,,
\end{align}
which is in agreement with \eqref{ffour}.

To summarize, given a deformation of an associative algebra, we can explicitly construct an $A_\infty$-algebra that can be thought of as a noncommutative deformation of this algebra, where the deformation parameter is promoted to an element of the algebra itself.\footnote{Let us mention another quite general approach to the deformation problem above. It is based on the construction of an appropriate resolution for the initial algebra. The approach is applicable to associative \cite{Sharapov:2018hnl,Sharapov:2017lxr} as well as to $A_\infty/L_\infty$-algebra deformations \cite{Sharapov:2018ioy,Li:2018rnc}. The choice of a resolution, however, is rather ambiguous and suitable resolutions may happen to be quite cumbersome. The advantage of the present approach is that it does not require any structure beyond the deformation  of the underlying associative algebra and, in this sense, it is more universal.} Remarkably, the $A_\infty$-structure is determined by the deformed product up to an $A_\infty$ gauge transformation. While the construction above is quite general, in the sequel we focus upon the case of  higher spin algebras and explain why and how these algebras can be deformed.

\section{Higher Spin Algebras}
\label{sec:HSA}
In the first approximation, higher spin algebras (HSA) are just (infinite-dimensional) associative algebras that arise in the study of higher symmetries of linear conformally invariant equations or of higher spin extensions of gravity. Very often the same algebras show up in other contexts under different names. For instance, one of the simplest examples is just the Weyl algebra $A_n$. Many examples of HSA are provided by various free conformal fields theories, being free they possess infinite-dimensional algebras of symmetries. Below we give a number of (almost) equivalent definitions and examples of HSA. The most important definitions for our subsequent discussion are due to free CFT's and universal enveloping algebras.

\subsection{Various Definitions and Constructions}

\paragraph{1. Higher symmetries of linear equations.} Given a linear equation $L\phi=0$, where $\phi\equiv \phi(x)$ is a set of fields and $L=L(x,\pl)$ is a differential operator, it is useful to study its symmetries and the algebra they form. A differential operator $S=S(x,\pl)$ is called a symmetry if it maps solutions to solutions, i.e., $LS\phi=0$ for any $\phi$ obeying $L\phi=0$. In practice, this implies that $L$ can be pushed through $S$, i.e., $LS=B_SL$ for some operator $B_S$. The operators of the form $CL$ are called trivial symmetries. These should be quotiented out as they act trivially on-shell. It is also important that the product $S_1S_2$ of two symmetries is a symmetry, as a consequence of linearity. Therefore, the algebra of symmetries -- the algebra of all symmetries modulo trivial ones -- is associative.

A canonical example \cite{Nikitin1991,Eastwood:2002su} is a free scalar field $\phi(x)$ in $d$-dimensional flat space and $L=\square$. The equation $\square \phi=0$ is well known to be conformally invariant, with conformal symmetries acting as\footnote{$a,b,c,\ldots =0,\ldots ,d-1$ are the indices of the Lorentz algebra $so(d-1,1)$.}
\begin{align}
    \delta_\xi \phi(x)&= \xi^a \pl_a \phi(x) +\frac{d-2}{2d}(\pl_a \xi^a) \phi(x)\,, && \pl^a\xi^b+\pl^b\xi^a=\frac{2}{d}\eta^{ab}\pl_m \xi^m\,,
\end{align}
where $\xi^a(x)$ is a conformal Killing vector. These symmetries form the conformal algebra $so(d,2)$ with respect to the commutator  $[\delta_{\xi_1},\delta_{\xi_2}]=\delta_{[\xi_1,\xi_2]}$. As is pointed out above, the product $\delta_{\xi_1}\cdots \delta_{\xi_n}$ is a symmetry too and is represented by a higher-order differential operator. All such operators are related to the conformal Killing tensors
\begin{align}
    \delta_v\phi&= v^{a_1...a_{k-1}}\pl_{a_1}\cdots \pl_{a_{k-1}}\phi+\text{more}\,, && \pl^{a_1}v^{a_2\ldots a_k}+\text{permutations}-\text{traces}=0\,.
\end{align}
It can be shown that the products of conformal symmetries generate the full symmetry algebra \cite{Nikitin1991,Eastwood:2002su}. Higher powers of Laplacian, $L=\square^k$, are also conformally-invariant operators with interesting symmetry algebras \cite{Gover, Bekaert:2013zya}. The symmetries of the free Dirac equation $\slashed \pl\psi=0$ \cite{Nikitin1991fer} and of many other relativistic wave equations are also known \cite{Pohjanpelto:2008st}.

The examples just given lead to infinite-dimensional associative algebras that contain the conformal algebra $so(d,2)$ as a (Lie) subalgebra under commutators. A possible generalization is to consider other (not necessarily conformally invariant) differential operators, e.g. massive Klein-Gordon equation.

To summarize, Definition 1: the higher spin algebras are defined to be the (associative) symmetry algebras of linear conformally-invariant equations.

\paragraph{2. Higher spin currents and charges.} Given a free field obeying $\square$-type equations, e.g. $\square \phi=0$, one can construct an infinite number of conserved tensors \cite{Craigie:1983fb,Anco:2002xn}
\begin{align}
    j_{a_1\ldots a_s}&= \phi \pl_{a_1}\ldots \pl_{a_s}\phi +\text{more terms}\,, && \pl^m j_{ma_2\ldots a_s}=0\,.
\end{align}
Due to the conformal invariance the conserved tensors can be made traceless and are thereby quasi-primary operators of the free boson CFT. Contracting them with conformal Killing tensors, one obtains conserved currents and the corresponding charges:
\begin{align}
    j_m(v)&= j_{ma_2\ldots a_s} v^{a_2\ldots a_s} \,,& Q_v&= \int d^{d-1}x \, j_0\,.
\end{align}
Definition 2 identifies higher spin algebras with the symmetries generated by the Noether charges associated to the higher spin currents. Via the Noether theorem Definition 2 is more or less equivalent to Definition 1. Such conserved tensors and symmetries associated to them  have been known since the 60's, see e.g. \cite{Deser:1980fk,Craigie:1983fb} and references therein. It was also shown that they do not survive when interactions are switched on. For CFT's the opposite statement is also true: the existence of conserved higher rank tensors implies that the theory is free in disguise  \cite{Maldacena:2011jn,Boulanger:2013zza,Alba:2013yda,Alba:2015upa}. The extensions of the Poincar{\'e} symmetry are constrained by the  Coleman--Mandula theorem \cite{Coleman:1967ad}.

\paragraph{3. Quotients of universal enveloping algebras.} A more direct description of HSA associated with linear conformally-invariant equations is via universal enveloping algebra $U(so(d,2))$, as the last paragraph  of item 1 suggests: juxtaposing conformal transformations generates the associative symmetry algebra. Therefore, we may collect the generators $P^a,K^a,L^{ab},D$ associated with the conformal algebra into $T^{AB}$ of $so(d,2)$.\footnote{$A,B,C,...=0,...,d+1$ are the indices of the conformal algebra $so(d,2)$ and $\eta^{AB}=(-+\cdots +-)$. Then, $L_{ab}=T_{ab}$, $D=-T_{d,d+1}$, $P_a=M_{a,d+1}-M_{a,d}$, $K_a=M_{a,d+1}+M_{a,d}$.} Then, any polynomial  
\begin{align}
    f(T^{AB})&= f(P^a,K^a,L^{ab},D)
\end{align}
generates a symmetry transformation. However, there are some relations meaning that not all the polynomials are independent and generate nontrivial transformations. For example, for the free scalar field we obviously have $P_aP^a\sim0$. The fundamental field corresponds to an irreducible representation of the conformal algebra and hence the Casimir operators have fixed numerical values. As a result, the symmetry algebra is isomorphic to the quotient of the universal enveloping algebra $U(so(d,2))$ by a two-sided ideal (annihilator) $\Jos$:
\begin{align}
    hs\sim U(so(d,2))/\Jos\,.
\end{align}
A concrete definition of $\Jos$ depends on a free CFT (irreducible representation) we consider, but, on general grounds, we expect all Casimir operators $\boldsymbol{C}_{2i}$ to have some fixed values $C_{2i}$. In the cases we are aware of $\Jos$ is generated by a few elements of $U(so(d,2))$. 

In the case of the smallest unitary representation, e.g. the free conformal scalar field, the annihilator $\Jos$ is also known as the Joseph ideal \cite{joseph1974}. Possible generalizations here is to consider more general ideals in $U(g)$ for any (not necessarily conformal) Lie algebra $g$, see e.g. \cite{Boulanger:2011se, Alkalaev:2014nsa, Joung:2015jza}. A useful for our studies example is provided by the HSA of the generalized free field CFT.\footnote{Note that Definitions 1 and 2 do not apply here, while the Definitions 3 and 4 can still be used, see below.} 

We stress that higher spin algebras depend on dimension $d$ and on the spectrum of the free CFT in a crucial way. Given a higher spin algebra one can read off the dimension of the spacetime, the spectrum of higher spin currents and the fundamental free field they are generated by. We conclude by Definition 3: the higher spin algebras are defined as various (simple) quotients of $U(so(d,2))$. 

\paragraph{4. Quantization of coadjoint orbits.} There is also a relation \cite{Eastwood:2002su,Fronsdal2009,Michel2014} between HSA and deformation quantization \cite{Fedosov:1994zz,Kontsevich:1997vb}. The fundamental field of any free CFT corresponds to some irreducible representation of the conformal algebra. This representation, in its turn, is associated to a certain coadjoint orbit (usually to a minimal nilpotent one). Not surprisingly that a given  HSA can be identified with the quantized algebra of functions on this coadjoint orbit. Possible generalizations here is to consider deformation quantization in full generality, i.e., for general symplectic or Poisson manifolds.

\subsection{Examples}
Let us discuss a few simple examples of HSA that will be important later. We mostly employ the universal enveloping realization of HSA. The conformal or anti-de Sitter algebra generators $T_{AB}$ obey 
\begin{align}
[T_{AB},T_{CD}]&=T_{AD}\eta_{BC}-T_{BD}\eta_{AC}-T_{AC}\eta_{BD}+T_{BC}\eta_{AD}\,,
\end{align}
and by the Poincar{\'e}--Birkhoff--Witt theorem, the decomposition of the universal enveloping algebra $U(so(d,2))$ is given by symmetrized tensor products of the adjoint representation\footnote{The language of Young diagrams is useful here. For example, the fundamental and the adjoint representations are depicted by \YoungmA{} and \YoungmAA, respectively. The trivial representation is denoted by $\bullet$.}
\begin{align}
    U(so(d,2))&=\bullet\oplus \YoungpAA \oplus \left[\,\YoungpBB\oplus \YoungpB\oplus\YoungpAAAA\oplus \bullet\right]\oplus \left[\,\YoungpCC\oplus\YoungpBBAA\oplus\cdots \right]\oplus\cdots \,.
\end{align}
Here the first singlet $\bullet$ is the unit of $U(so(d,2))$, $\YoungmAA\sim T^{AB}$ and the second $\bullet$ is the quadratic Casimir operator
\begin{align}
    \boldsymbol{C}_2=-\frac12 T_{AB} T^{AB}\,.
\end{align}
In what follows we describe some ideals of $U(so(d,2))$ and the corresponding quotients that yield the HSA of interest. 

\paragraph{Free Boson HSA.} This is the simplest HSA and the generators of the ideal can be guessed from the symmetries of $\square \phi=0$. Since the solution space is an irreducible representation, the values of the Casimir operators are fixed. Decoupling of null states implies $P_aP^a=0$ and $K^a K_a=0$. Finally, all anti-symmetric combinations of the conformal symmetry generators, e.g. $L_{[ab}P_{c]}$ and $L_{[ab}L_{cd]}$, should vanish. All in all, the two-sided (Joseph) ideal is generated by \cite{Eastwood:2002su}\footnote{The full two-sided ideal is obtained by taking the generators and multiplying them by $U(so(d,2))$.}  
\begin{align}\label{EastwoodI}
    \Jos&=\YoungpAAAA\oplus \YoungpB\oplus \left(\boldsymbol{C}_2 -C_2\right)\,, &
    &C_2=-\frac{1}{4}(d^2-4)\,.
\end{align}
The $so(d,2)$ decomposition of the quotient algebra contains traceless tensors described by  rectangular, two-row, Young diagrams:
\begin{align}\label{EastwoodHS}
    hs_{F.B.}&=\bullet \oplus \YoungpAA\oplus \YoungpBB\oplus \YoungpCC\oplus\cdots\,.
\end{align}
More explicitly, the generators of the Joseph ideal read:
\besubeqs
\begin{align}
    \Jos^{ABCD}&= T^{[AB} T^{CD]}\,,\\
    \Jos^{AB}&= T\fud{A}{C} T^{BC}+T\fud{B}{C} T^{AC}-(d-2)\eta^{AB}\,,\\
    \Jos&= -\frac12 T_{AB} T^{AB}+\frac14(d^2-4)\,.
\end{align}
\esubeqs
\paragraph{Free Boson and Free Fermion HSA in Three Dimensions.} This is an even simpler example since all of the Joseph ideal relations can be resolved thanks to the isomorphism $so(3,2)\sim sp(4)$.\footnote{Some important facts are contained already in \cite{Dirac:1963ta}. Everything we discuss below can be found in \cite{Gunaydin:1981yq,Gunaydin:1983yj,Gunaydin:1989um}.} It turns out that the free boson and free fermion fields -- as representations of $sp(4)$ -- are equivalent to even and odd states in the Fock space of the $2d$ harmonic oscillator:
\besubeqs\label{3donshell}
\begin{align}
    P^{a_1}...P^{a_k}|\phi\rangle &\sim a^\dag_{\ga_1}...a^\dag_{\ga_{2k}} |0\rangle\,,\\
    P^{a_1}...P^{a_k}|\psi\rangle_\gd &\sim a^\dag_{\ga_1}...a^\dag_{\ga_{2k}} a^\dag_\gd|0\rangle  \,. 
\end{align}
\esubeqs
Here $a^\ga$ and $a^\dag_\gb$  are the standard creation/annihilation operators satisfying 
\begin{equation}
[a^\ga,a^\dag_\gb]=\delta^\ga_\gb\,,\qquad a^\ga|0\rangle=0\,,
\end{equation}
and $\ga,\gb=1,2$ are the spinor indices from the $so(1,2)$ point of view. The spinor-vector dictionary is through the $\sigma$-matrices, e.g. $P_m=\sigma_m^{\ga\gb}a^\dag_\ga a^\dag_\gb$. The $sp(4)$ generators are realized by the ten bilinears in $a^\ga$ and $a^\dag_\ga$:
\begin{align}\label{conformalbase}
    K^{\ga\gb}&=a^\ga a^\gb\,, & \frac12D\delta^\ga_\gb+L^\ga_\gb&=\frac12 \{a^\ga,a^\dag_\gb\}\,, & P_{\ga\gb}&=a^\dag_\ga a^\dag_\gb  \,.
\end{align}
This is the standard oscillator realization of $sp(4)$. The algebra of all ordered polynomials $O(a,a^\dag)$ in $a^\ga$, $a^\dag_\gb$ is the Weyl algebra $A_2$.\footnote{The subscript indicates the number of canonical pairs, two in our case.} The HSA, as an algebra that maps on-shell states \eqref{3donshell} to the on-shell states, is the even subalgebra of the Weyl algebra $A_2$, i.e., $O(a,a^\dag)=O(-a,-a^\dag)$. 

The most important feature of the $3d$ case is that the HSA's of free boson and free fermion fields are equivalent and isomorphic to the even subalgebra of $A_2$. This is not true when $d>3$ for an obvious reason that the higher spin currents built out of the free fermion do not match those of the free boson, see e.g. \cite{Alkalaev:2012rg}. 

\paragraph{Generalized Free Field HSA.} A generalized free (scalar) field, i.e., a conformal scalar operator $O_\Delta(x)$ of some weight $\Delta$ such that all correlators are computed via the free Wick contractions,\footnote{For generic $\Delta$, the generalized free field does not have a local stress-tensor and does not have (local) higher spin currents. Also, there are no equations to be imposed. Therefore, the definitions (1) and (2) are not applicable. Nevertheless, the algebra can be defined via definition (3) (and also via (4)) as we do here. A good consistency check is that it reduces to the already known HSA at the expected values of $\Delta$. } is a useful approximation in many situations. The corresponding HSA, denoted by $hs_\Delta$, is defined to be the quotient $hs_\Delta =U(so(d,2))/\Jos_\Delta$ with respect to the ideal generated by
\begin{align}\label{GFFI}
    \Jos_\Delta&=\YoungpAAAA\oplus \left(\boldsymbol{C}_2 -C_2(\Delta)\right)\,, &
    &C_2(\Delta)=\Delta(d-\Delta)\,,
\end{align}
or, in components,
\begin{align}\label{JGFF}
    \Jos^{ABCD}&= T^{[AB} T^{CD]}\,, & 
    \Jos&= -\frac12 T_{AB} T^{AB}-C_2(\Delta)\,.
\end{align}
The interpretation of the ideal is obvious. That $\Jos^{ABCD}$ must vanish is manifestation of the lowest state  $|\Delta\rangle$ being scalar, which implies that the descendants $P^a\ldots P^c|\Delta\rangle$ are symmetric tensors and the combinations of the generators with more than two anti-symmetrized indices vanish. The $so(d,2)$-decomposition contains more tensors than that of the free boson HSA, namely,  
\begin{align}
    hs_\Delta&=\bullet \oplus \YoungpAA\oplus \YoungpBB\oplus \YoungpB\oplus \YoungpCC\oplus\YoungpCA\oplus\cdots\,.
\end{align}
The additional components are due to the absence of the $\YoungpB$ generator, c.f. \eqref{EastwoodI} and \eqref{GFFI}.\footnote{It may seem that one can pick several elements of $U(so(d,2))$ in random and declare them to generate an ideal, but in doing so one may discover that the ideal coincides with the full $U(so(d,2))$. In particular, it is impossible to add the $\YoungpB$ component to the generating set for generic $\Delta$ without trivializing the quotient.}

Clearly, the HSA of generalized free field $O_\Delta(x)$ form a one-parameter family of algebras because $\Delta$ is a free parameter. At the critical values $\Delta_k=d/2-k$, $k=1,2,\ldots$, the algebra is not simple and acquires a two-sided ideal. The resulting quotient algebra is the symmetry algebra of the free scalar field   $\square^k \phi=0$ \cite{Gover, Bekaert:2013zya}. The one-parameter family of HSA corresponding to generalized free fields will be important for the discussion in Appendix \ref{sec:HSAdeformed} since it underlies the deformation of the other HSA.

\subsection{Higher Spin Currents Equal Higher Spin Algebra}
\label{sec:HScurrents}
As it was already mentioned, the higher spin symmetry of free CFT's is manifested by an infinite number of higher-spin currents $J_s$, which are quasi-primary operators from the CFT point of view. Schematically, e.g. in the free scalar CFT, they are
\begin{align}
   J_{a_1\ldots a_s}&= \phi \pl_{a_1}\ldots \pl_{a_s}\phi +\text{more}\,, && \pl^cJ_{ca_2\ldots a_s}=0 \,.
\end{align}
The stress-tensor, which is responsible for the $so(d,2)$-part of the HSA is the $s=2$ member of the family. By construction, the free field is a fundamental representation of this HSA.\footnote{Representations (modules) of HSA are quite easy to describe, see e.g. \cite{Beisert:2004di}. Roughly speaking, the free field is a vector space $V$ and HSA is $gl(V)$ for this $V$. Other representations are just tensor products $V\otimes \cdots \otimes V$ projected onto any irreducible representation of the permutation group (the permutation group commutes with the $gl(V)$-action on $T(V)$).} The infinite multiplet $J$ of higher spin currents $J_s$ is the representation that is next to the fundamental one.\footnote{One should be careful about tensor product vs. associativity issues and imply either the Lie subalgebra of a HSA (via commutators) or the tensor product of HSA that naturally acts on the tensor product of its representations. } The lowest lying OPE's can be written as
\begin{align}
    \phi \phi&= \mathds{1}+J\,, && JJ=\mathds{1}+J+O_2    \,,
\end{align}
where $\mathds{1}$ is the identity operator and $O_2$ is a multiplet of double-trace operators, which is given by the quartic tensor product of the free field itself. 

Regarding  the free field as a vector space $V$ and HSA as $gl(V)$, the higher spin currents belong to $V\otimes V$, which is very close to $gl(V)\sim V\otimes V^*$. This heuristic reasoning can be made more precise.\footnote{See \cite{Basile:2018dzi} for subtleties that may arise in some formal manipulations. That the tensor product decomposes into (all) higher spin currents was shown, for $d=3$, in \cite{Flato:1978qz} (the currents, as representations of $so(d,2)$, viewed as anti-de Sitter algebra, are the same as massless fields in $AdS_{d+1}$, which is the interpretation adopted in \cite{Flato:1978qz}). See \cite{Craigie:1983fb} for the result in any $d$. See also \cite{Iazeolla:2008ix} that elaborates on the relation between this construction and $U(so(d,2))$, showing, in particular, that the shadow of $J_0$ can also be treated by the same tools. } If $|\phi\rangle$ is the free field vacuum, then 
\begin{equation}
K^a|\phi\rangle=0\,,\qquad L^{ab}|\phi\rangle=0\,,\qquad D|\phi\rangle=\tfrac{d-2}2|\phi\rangle
\end{equation} 
and the descendants correspond to $P^a\ldots P^c|\phi\rangle$. Higher spin currents are the quasi-primary states in the tensor product
\begin{align}
   J\sim  \phi \times \phi \sim P^a\ldots P^c|\phi\rangle\otimes P^b\ldots P^d|\phi\rangle\,,
\end{align}
while the HSA can be viewed as the span of operators of the form
\begin{align}
     P^a\ldots P^c|\phi\rangle\otimes \langle\phi|K^b\ldots K^d\,.
\end{align}
Clearly, the two spaces are formally isomorphic and the map between them is the conjugation $\langle\phi|=|\phi\rangle^\dag$, which is defined via the inversion map $\Imap$.\footnote{Note that at the level of the Lie algebra we have $K^a=IP^aI$, $L^{ab}=IL^{ab}I$, $P^a=IK^aI$ and $-D=IDI$. We see that $P^a+K^a$ and $L^{ab}$ are stable and form $so(d-1,2)$ subalgebra of the conformal algebra $so(d,2)$. We can also define $-K^a=IP^aI$, $L^{ab}=IL^{ab}I$, $-P^a=IK^aI$ and $-D=IDI$. Then, it is  $P^a-K^a$ and $L^{ab}$ that are stable and form $so(d,1)$.} Therefore, the higher spin currents together with their descendants, as a module of the conformal algebra (as well as a HSA-module), can be viewed as the same HSA where the right action is twisted by $\Imap$. That is, $J\Imap$ is formally isomorphic to HSA.  

\section{Slightly Broken Higher Spin Symmetry}
\label{sec:SlightlyBroken}
Now we are ready to explain the problem of the slightly broken higher spin symmetry and our proposal in greater detail. In interacting CFT's with slightly broken higher spin symmetry higher spin currents are no longer conserved, but their non-conservation has a very specific form of 
\begin{align}\label{noncons}
    \pl \cdot J&= \frac{1}{N} [JJ]\,,
\end{align}
where $[JJ]$ is a specific (set of) double-trace operators, whose form may also depend on the coupling constants (e.g. it depends on $\lambda=N/k$ for Chern--Simons matter theories), see \cite{Maldacena:2012sf,Skvortsov:2015pea,Giombi:2016hkj,Giombi:2016zwa,Giombi:2017rhm} for some explicit formulas. It is worth mentioning at this point, that the non-conservation equations for the two theories related by the bosonization duality can be directly mapped into each other \cite{Giombi:2016zwa}, which again supports the statement that \eqref{noncons} and its consequences should explain the dualities.\footnote{We also note that there are cases where the triple-trace terms $N^{-2}[JJJ]$ are possible. For example, this corresponds to the sextic coupling $\lambda_6(\phi^2)^3$ in the action. The fate of such terms is yet unclear to us. First of all, $\lambda_6$ is not an independent parameter since the conformal point corresponds to $\beta_{\lambda_6}=0$ \cite{Aharony:2018pjn}. Secondly, the triple-trace terms are suppressed by an additional $N^{-1}$ and, for example, play no role for the anomalous dimensions and three-point functions studied in \cite{Giombi:2016zwa}. The same time, such triple terms may be (already) accounted for by the higher structure maps of the $A_\infty$-algebra. }

At the free level a given CFT, including the correlation functions, is fully controlled by the corresponding higher spin algebra (HSA), say $hs$. The multiplet of higher spin currents $J$ form a representation of $hs$. Moreover, $J\Imap$ is formally isomorphic to $hs$ (as representation). The slightly broken higher spin symmetry is a deformation of $hs$ by $J$ (to be precise by $J/N$ to introduce a small parameter). Since $J\Imap \sim hs$, we can interpret the sought for deformation as a non-commutative deformation of $hs$ described in Section \ref{sec:Aconstruction}. Then, the HSA belongs to $A_{-1}$ and the non-commutative deformation parameter to $A_{0}$, i.e., $A_{-1}\sim hs$ and $A_0\sim J$. The only subtlety is that one should take into account the inversion map $\Imap$. From the $A_\infty$ point of view the action of a HSA on the module should be defined as
\begin{align}\label{taction}
    m_2(a,u)&=au\,, &&m_2(u,a)=-u\Imap(a) \quad \text{for}\quad a\in A_{-1}\,, u\in A_{0}\,.
\end{align}
We show below that this problem can be reduced to the one already solved in Section \ref{sec:Aconstruction}. The procedure is three step. Firstly, we extend $hs$ by adding the automorphism $\Imap$ and call the resulting algebra the double $D(hs)$. Secondly, we deform $D(hs)$ as an associative algebra. This deformation can be used to construct an appropriate  $A_\infty$-algebra with the help of Section \ref{sec:Aconstruction}. Lastly, we can truncate the algebra in such a way that \eqref{taction} is true.

As it was already mentioned, typical  HSA's admit no deformations as associative algebras, which means that $HH^2(hs,hs)=0$.\footnote{What we discuss below applies also to the examples where they do admit such deformations.} Nevertheless, certain simple extensions of HSA's do admit deformations and it is these deformations that are also responsible for the $A_\infty$-structure.

In order to treat both the HSA $hs$ and $J$ on an equal footing, we take a bigger algebra --- the double $D(hs)$ --- HSA extended by $\Imap$. This is just the simplest example of the smash product $B\rtimes \Gamma$, where $B$ is an algebra and $\Gamma$ is a finite group of its automorphisms of $B$. In our case $\Gamma=\mathbb{Z}_2$. Elements of $D(hs)$ have the form $a=a'+a''\Imap$, $a',a''\in hs$ and the product law reads 
\begin{equation}
(a'+a''\Imap)(b'+b''\Imap)=(a'b'+a'' \Imap(b'')) +(a'b''+a'' \Imap(b'))\Imap\,,
\end{equation} 
where $\Imap(a)$ is the action of the inversion on the algebra elements, which can be obtained by extending $\Imap(P^a)=\Imap P^a\Imap=K^a$, etc. to polynomials in $P^a,K^a, L^{ab}, D$ and we used $\Imap^2=1$. Now, the usual adjoint action and the twisted action \eqref{taction} are just different projections of the adjoint action in $D(hs)$. 

An important observation is that $D(hs)$ belongs to a one-parameter family of algebras (while $hs$ usually does not). We discuss various arguments in favor of this statement in Appendix \ref{sec:HSAdeformed} and explicit examples in Section \ref{sec:Oscillators}. For now it is sufficient to assume that we have already constructed a one-parameter family of associative algebras $D_\hbar(hs)$ that deforms the double $D(hs)$. As a result one gets the deformed product \eqref{absdeformed}:
\begin{align}\label{absdeformedB}
    a\ast b&= ab+\sum_{k>0}\phi_k(a,b)\hbar^k && a,b\in D(hs)\,.
\end{align}
We also assume that such deformation does not originate from a deformation of $hs$ itself in those exceptional cases when the latter exists.\footnote{It is worth stressing that even if a given $hs$ happens to belong to a one-parameter family of algebras, it will not lead to the $A_\infty$-algebra we need, for the $m_2$-map has to be \eqref{taction} to incorporate higher spin currents. } This means that the first-order deformation has to obey
\begin{align} \label{Ideform}
    a\phi_1(b,c)-\phi_1(ab,c)+\phi_1(a,bc)-\phi_1(a,b)I(c)&=0\,, &&a,b,c\in hs\,.
\end{align}
The same equation without $\Imap$ determines the first-order deformations of $hs$. A nontrivial solution to \eqref{Ideform} is a Hochschild cocycle in the representation twisted by $\Imap$. Then, the first-order deformation of $D(hs)$ induced by $\phi_1$ can be written as
\begin{align*}
    (a+a'I)\ast (b+b'I)=(ab+aI(b'))+(ab'+a'I(b))I+\hbar\phi_1(a+a'I,b+b'I)I+\mathcal{O}(\hbar^2)\,.
\end{align*}
This illustrates the relation between the Hochschild cohomology in the representation twisted by $\Imap$, \eqref{Ideform}, and the Hochschild cohomology of the double $D(hs)$. 

The $A_\infty$-algebra is constructed by building up the structure maps $m_n$ following the general method of Section \ref{sec:Aconstruction}. Since $\Imap^2=1$ the Taylor coefficients $\phi_k$ have a specific dependence on $\Imap$: $\phi_{2n+1}(a,b)=\varphi_{2n+1}(a,b)\Imap$ and $\phi_{2n}(a,b)=\varphi_{2n}(a,b)$ where $\varphi_k$ do not depend on $\Imap$. This property makes it obvious that we can restrict $A_{-1}$ to $hs$, while all elements from $A_{0}$ can be restricted to $hs \Imap$ to be interpreted as $J\Imap$ for the multiplet of higher spin currents $J$. Assuming that $a,b$ and $u,v$ take values in $hs$, we can write for the first structure maps $m_n$
\begin{eqnarray*}
    &m_2(a,b)= ab\,, \qquad m_2(a,u)=au\,, \qquad  m_2(u,a)=-u\Imap(a)\\
    &m_3(a,b,u)=\varphi_1(a,b)\Imap(u) \,, \qquad m_3(a,u,v)=\varphi_1(a,u)\Imap(v)\,, \qquad
    m_3(u,a,v)=-\varphi_1(u,\Imap(a))v \,,\\
    &m_4(a,b,u,v)=\varphi_2(a,b)u\Imap(v)+\varphi_1(\varphi_1(a,b), \Imap(u))\Imap(v)\,, \qquad \ldots
\end{eqnarray*}
The bilinear structure maps $m_2$ are equivalent to having a higher spin algebra and its module in the adjoint representation twisted by $\Imap$, i.e.,  higher spin currents.  
One can check that these structure maps do obey the first defining relations of the $A_\infty$-algebra. The general formula for $m_n$ is in Section \ref{sec:Aconstruction}. Further investigation requires explicit form of $\phi_k$ and some examples are provided in Section \ref{sec:Oscillators}.

\section{Explicit Oscillator Realizations}
\label{sec:Oscillators}
In practice we may need an efficient way to perform computations with the deformed algebras. Many interesting HSA's admit oscillator realizations and, after briefly reviewing these realizations, we modify them as to construct the deformed HSA's. Note that the structure maps of the $A_\infty$-algebra are expressed in terms of the deformed product and as such contain no new information compared to the one in the deformed HSA. Therefore, the problem of the slightly broken higher spin symmetry is reduced to a much simpler problem of constructing a deformed HSA. This brings up the question: How big is the space of all deformations?  
Quantitatively its size is defined by the second Hochschild cohomology group (if there are no obstructions), whose dimension equals the number of  phenomenological parameters entering the correlation functions.

\subsection{Toy Model: Weyl Algebra \texorpdfstring{$A_1$}{}}
The simplest example, which nevertheless underlies all the other deformations, is the smallest Weyl algebra $A_1$, i.e., a one-dimensional harmonic oscillator. The Weyl algebra $A_1$ is defined in our notation as\footnote{Here, $\epsilon_{\ga\gb}$ is the invariant $sp(2)$-tensor, the anti-symmetric tensor with $\epsilon_{12}=-\epsilon_{21}=1$.}
    \begin{align}
     [y_\ga,y_\gb]&=2i \epsilon_{\ga\gb}\,, && \ga,\gb=1,2\,.
    \end{align} 
Let us define the automorphism $\Imap $ as the reflection $\Imap (y_\ga)=-y_\ga$. Therefore, the $\Imap $-stable subalgebra -- the `Lorentz' subalgebra -- is simply the subalgebra $A_1^e$ of even polynomials in $y$'s, $f(y)=f(-y)$. It is well known that the Weyl algebra does not admit any deformation as an associative algebra, but the `Lorentz' subalgebra does belong to a one-parameter family of algebras. Indeed, $sp(2)\sim sl(2)$ is a subalgebra of the Weyl algebra, which is realized by the three generators $t_{\ga\gb}=t_{\gb\ga}$:
\begin{align}
   t_{\ga\gb}&=-\tfrac{i}{4}\{y_\ga,y_\gb\}\,,  &[t_{\ga\gb}, t_{\gc\gd}]&= \epsilon_{\ga\gd} t_{\gb\gc} +\text{three more}\,.
\end{align}
The $\Imap $-stable subalgebra $A_1^e$ coincides with the enveloping algebra of $t_{\ga\gb}$ and it is not hard to see that this is the quotient of $U(sl_2)$ by the two-sided ideal generated by $\boldsymbol{C}_2-(-\tfrac34)$, where $\boldsymbol{C}_2=-\tfrac12 t_{\ga\gb} t^{\ga\gb}$ is the Casimir operator; the constant $-\tfrac34$ is the value of $\boldsymbol{C}_2$ in the oscillator realization. This algebra belongs to a one-parameter family of algebras,\footnote{These algebras were defined in \cite{Feigin} and dubbed $gl_\lambda$ since they reduce to $gl_N$ for certain values of $\lambda$ and can be thought of as algebras interpolating between $gl_N$ and $gl_{N+1}$.} called $hs(\lambda)$ that are obtained in the same way except that the eigen value of the Casimir operator is kept to be a free parameter: 
    \begin{align}
        hs(\lambda)&=U(sl_2)/\Jos\,, && \Jos=U(sl_2)[\boldsymbol{C}_2+(\lambda^2-1)]\,.
    \end{align}
$hs(\lambda)$ is nothing but a noncommutative (fuzzy) sphere, whose radius is controlled by $\lambda$.   
Therefore, we have $A_1^e\sim hs(\lambda^*)$, where $\lambda^*=1/2$. According to our general claim, the Weyl algebra $A_1$ extended by the automorphism $\Imap $ should admit a one-parameter family of deformations. The double $D(A_1)$ is defined by
    \begin{align}
     [y_\ga,y_\gb]&=2i \epsilon_{\ga\gb}\,, &&\{y_\ga,k\}=0\,,  && k^2=1\,.
    \end{align} 
Indeed, the algebra generated by the $y$'s and $k$ is a particular case of the so-called deformed oscillator algebra $Aq(\nu)$,\footnote{Defined implicitly in \cite{Wigner} and explicitly in e.g. \cite{Yang:1951pyq,Mukunda:1980fv,Ohnuki:1985wg}, see also \cite{Vasiliev:1989re}.} which is defined by the following relations on its generators:
         \begin{align}
         [q_\ga,q_\gb]&=2i \epsilon_{\ga\gb}(1+\nu K)\,, &&\{q_\ga,K\}=0 && K^2=1\,.
        \end{align}   
It is clear that the double of the Weyl algebra $D(A_1)$ is isomorphic to $Aq(0)$. Another description of the deformed oscillator algebra is
    \begin{align}
        Aq(\nu)&= U(osp(1|2))/\Jos\,, && \Jos=U(osp(1|2))[C_2+\frac14(1-\nu^2)]\,.
    \end{align}
This algebra is nothing but a noncommutative super-sphere $S^{2|2}$ whose radius is controlled by $\nu$. The structure constants of $hs(\lambda)$ and of the deformed oscillators are available in the literature in several forms \cite{Pope:1989sr, Korybut:2014jza, Fradkin:1990qk, Joung:2014qya, Basile:2016goq}. Therefore,  the components $\phi_k(\bullet,\bullet)$ of the deformed HSA product are known and can be used to explicitly write down the $A_\infty$-structure. Notice that the classical limit of the deformed algebra is just a two-dimensional symplectic space endowed  with a symplectic reflection $k$.

One may wonder to which extent the deformation described above is unique. For the Weyl algebra it is well known that $HH^2(A,A^*)$ is one-dimensional. At the same time, the $\Imap $-map identifies the dual module $A^*$ with the $\Imap $-twisted one. For the double $D(A_1)$ the cohomology is known to be one-dimensional and the deformation is unique.

\subsection{Deformations of the Free Boson Algebra}
The simplest example of a HSA is the symmetry algebra of the free boson CFT \cite{Eastwood:2002su}. The case of three dimensions is somewhat special and is discussed in the next section. The $A_\infty$-algebra originating from this HSA should be responsible for the breaking of higher spin symmetries in the large-$N$ critical vector model in $d$ dimensions.\footnote{Due to the unitarity constraints the unitary cases are confined to $2<d<4$ and $4<d<6$ \cite{Fei:2014yja}. It would be interesting to extend the $A_\infty$-algebra to fractional dimensions $d$.}

There exists a quasi-conformal realization of this HSA by the minimal number of oscillators where the Joseph ideal is completely resolved \cite{Fernando:2015tiu}. This realization is non-linear and for simplicity let us stick to another, linear, form \cite{Vasiliev:2003ev}, in which the Joseph ideal is partially resolved. Such a realization appears naturally in the manifestly conformally-invariant description of the free conformal scalar field in the ambient space \cite{Bekaert:2009fg}. One begins with the embedding of the HSA into the Weyl algebra $A_{d+2}$:\footnote{Here $A,B,\ldots =0,\ldots ,d+1$ are indices of $so(d,2)$. We will also split them as $A=\{\aA,5\}$, etc., where $\aA,\aB,\ldots =0,\ldots ,d$ are the indices of the AdS-Lorentz algebra $so(d,1)$ and $5$ is an extra dimension. $L^{\aA\aB}=T^{\aA\aB}$, $P^\aA=T^{\aA5}$, $\eta^{55}=-1$, so that $[P^\aA,P^\aB]=L^{\aA\aB}$.}
\begin{align}
[Y^A_\ga,Y^B_\gb]&=2i\eta^{AB}\epsilon_{\ga\gb} \,.
\end{align}
The bilinears in $Y$ form $sp(2(d+2))$, which contains a Howe dual pair $so(d,2)\oplus sp(2)$ of algebras such that the $so(d,2)$ generators $T^{AB}$ commute with the $sp(2)$ generators $t_{\ga\gb}$:
\begin{align}
T^{AB}  &= +\frac{i}{4}\epsilon^{\alpha\beta}\{Y^A_\alpha, Y^B_\beta\}\,, &t_{\ga\gb}&=-\frac{i}4\{Y^A_\ga,Y_{A\gb}\}\,.
\end{align}
We consider the enveloping algebra of $T^{AB}$, i.e., polynomials $f(Y)\equiv f(T)$, which can also be defined as the centralizer of $sp(2)$,  $[t_{\ga\gb}, f(Y)]=0$. 
By construction, a part of the Joseph ideal vanishes identically since one cannot have more than two anti-symmetrized indices of $so(d,2)$:
\begin{align}
    T^{[AB}T^{CD]}\sim \YoungmAAAA\sim0\,.
\end{align}
The resulting algebra is not simple and its $so(d,2)$ decomposition contains traceful tensors with the symmetry of two-row rectangular Young diagrams:
\begin{align}
    f(T)\sim \bullet \oplus \YoungSLAA\oplus \YoungSLBB\oplus \YoungSLCC\oplus\cdots \,.
\end{align}
The HSA is defined as a quotient of this algebra by the ideal generated by traces:
\begin{align}\label{subquotient}
f\in hs_{F.B.}&: && [t_{\ga\gb},f]=0\,, && f\sim f+ t_{\ga\gb}\star g^{\ga\gb}\,,
\end{align}
where $g^{\ga\gb}$ transforms as an $sp(2)$-tensor. Note that the $sp(2)$-generators $t_{\ga\gb}$ are exactly the contractions of $Y$'s, that is, traces. The resulting spectrum is \eqref{EastwoodHS}, as expected. 

The automorphism $\Imap $ that corresponds to the inversion map in the CFT base and to the flip of the AdS-translations in the AdS base is realized as $\Imap (y^\aA_\ga,y_\ga)=(y^\aA_\ga,-y_\ga)$, i.e., it flips the sign of the $A_1$ subalgebra generators. Since the $\Imap $-map does not affect $y^\aA_\ga$, the whole construction is very similar to the $A_1$ toy model. The double $D(hs_{F.B.})$ is easy to construct:
\begin{align}
    [y^\aA_\ga, y^\aB_\gb]&= +2i \epsilon_{\ga\gb}\eta^{\aA\aB}\,, &
    [y_\ga,y_\gb]&=-2i\epsilon_{\ga\gb}\,,  && \{y_\ga,k\}=0\,.
\end{align}
The deformed double is then obtained with the help of the deformed oscillators,\footnote{Note that the inversion map can also be realized as $\Imap (y^\aA_\ga,y_\ga)=(-y^\aA_\ga,y_\ga)$, but this realization does not admit the deformation we are looking for. It is important to note that while the double $D(hs)$ can always be deformed, a particular (oscillator) realization of $D(hs)$ may not admit any straightforward deformation. Indeed, in the present case $hs$ is realized as a subquotient of the Weyl algebra $A_{d+2}$ and we are to deform $D(A_{d+2})$ first, which may or may not be possible depending on how $D(A_{d+2})$ is realized.  }
\begin{align}
    [y^\aA_\ga, y^\aB_\gb]&= +2i \epsilon_{\ga\gb}\eta^{\aA\aB}\,, &
    [q_\ga,q_\gb]&=-2i \epsilon_{\ga\gb}(1+\nu k)\,, && \{q_\ga,k\}=0\,,
\end{align}
and is defined following \eqref{subquotient} as
\begin{align}
  D_\nu(hs)\ni f(y^\aA_\ga,q_\ga,k)&:  &[f,t_{\ga\gb}]&=0\,, && f\sim f+t_{\ga\gb}\star g^{\ga\gb}(y,q,k) \,,
\end{align}
where the new $sp(2)$ generators are
\begin{align}
    t_{\ga\gb}&=-\frac{i}{4}\{y^\aA_\ga,y_{\aA\gb}\}+\tau_{\ga\gb}\,, &&
    \tau_{\ga\gb}=\frac{i}{4}\{q_\ga,q_{\gb}\}\,.
\end{align}
At this point, there is no need in the deformed oscillators themselves, it is sufficient to know that the deformation of the algebra in $y_\ga$ and $k$ is given by the quotient of $U(osp(1|2))$, the fuzzy super-sphere.

The first few levels of the deformed double are easy to explore. Following the general logic, one can define the Lorentz and translation generators
\begin{align}
    \Pd^\aA&=+\frac{i}{4} \{y^\aA_\ga, q_\gb\}\epsilon^{\ga\gb}\,, & \Ld^{\aA\aB}&=+\frac{i}{4} \{y^\aA_\ga, y^{\aB}_\gb\}\epsilon^{\ga\gb}
\end{align}
that commute with $sp(2)$: 
\begin{align}
    [t_{\ga\gb}, \Pd_\aA]&=0\,, & [t_{\ga\gb}, \Ld_{\aA\aB}]&=0\,,& [t_{\ga\gb}, k]&=0\,.
\end{align}
The relations of the $so(d,2)$ algebra get modified at one place
\begin{align*}
    [\Pd^\aA,\Pd^\aB]&= (1+\nu k)\Ld^{\aA\aB}\,, && [\Ld^{\aA\aB},\Ld^{\aC\aD}]=\Ld^{\aA\aD}\eta^{\aB\aC}+\ldots\,, &
    [\Ld^{\aA\aB},\Pd^\aC]&= \Pd^\aA\eta^{\aB\aC} -\Pd^\aB\eta^{\aA\aC}\,,
\end{align*}
which is the first nontrivial component of the Hochschild cocycle.

\subsection{Three Dimensions}
The case of three dimensions is special due to the fact that the HSA of the free boson CFT is the same as the HSA of the free fermion CFT. A unique HSA is the even subalgebra $A_2^e$ of the Weyl algebra $A_2$:\footnote{$A,B,...=1,...,4$ are the $sp(4)$ vector indices, $sp(4)\sim so(3,2)$. The $AdS$-Lorentz algebra is $sl(2,\mathbb{C})\sim so(3,1)$ and it is convenient to use the indices $\ga,\gb,\ldots=1,2$ and $\gad,\gbd,\ldots=1,2$ of the fundamental of $sl(2,\mathbb{C})$ and its conjugate. }
\begin{align}
hs\ni f(Y)&: && f(Y)=f(-Y)\,, & [Y^A,Y^B]&= 2iC^{AB}\,.
\end{align}
In the AdS-base the quartet $Y^A$ can be split into the commuting $y_\ga, \bar{y}_\gad$ in terms of which  
\begin{align}
    \Ld_{\ga\gb}&=-\frac{i}{4}\{y_\ga,y_\gb\}\,, &
    \Pd_{\ga\gad}&=-\frac{i}{4}\{y_\ga,\bry_\gad\}\,, &
    \brLd_{\gad\gbd}&=-\frac{i}{4}\{\bry_\gad,\bry_\gbd\}\,.    
\end{align}
In the conformal base we have \eqref{conformalbase}. The $\Imap$-map acts either as $\Imap(y_\ga,\bry_\gad)=(-y_\ga,\bry_\gad)$ or as $\Imap(y_\ga,\bry_\gad)=(y_\ga,-\bry_\gad)$. In the conformal base it corresponds to $\Imap(a^\ga,a^\dag_\gb)=(a^\dag_\ga,a^\gb)$ or $\Imap(a^\ga,a^\dag_\gb)=(-a^\dag_\ga,-a^\gb)$. That there are two different realizations of the $\Imap$-map is in accordance with the existence of two independent cocycles, which was already deduced in Appendix \ref{sec:HSAdeformed} from the dual cycles.  
The double of this algebra is just the two copies of the one for $A_1$:\footnote{The same algebra appeared in \cite{Vasiliev:1986qx} as $\mathcal{N}=2$ super-HSA. }
\begin{align}
    \{y_\ga, k\}&=0\,, & [\bry_\gad,k]&=0\,, & \{\bry_\gad,\brk\}&=0\,, &[y_\ga,\brk]&=0\,.
\end{align}
The exact deformation of the double is given by the two pairs of deformed oscillators:
\begin{align}
[q_\ga,q_\gb]&=2i \epsilon_{\ga\gb}(1+\mu k)\,, &&\{q_\ga,k\}=0\,, && [\brq_\gad,k]=0\,,\\
[\brq_\gad,\brq_\gbd]&=2i \epsilon_{\gad\gbd}(1+\bar\mu \brk)\,, &&\{\brq_\gad,\brk\}=0\,, && [q_\ga,\brk]=0\,.
\end{align}  
The reality conditions $q^\dag_\ga= \brq_\gad$ imply that $\mu=\nu e^{i\theta}$, $\bar\mu=\nu e^{-i\theta}$ for real $\nu$. Geometrically, the deformed double of the HSA corresponds to the direct product of two noncommutative super-spheres $S^{2|2}\times S^{2|2}$ that have the same (absolute) value of radii. 

The deformed Lorentz and translation generators are given by the same formulae
\begin{align}
    \Ld_{\ga\gb}&=-\frac{i}{4}\{q_\ga,q_\gb\}\,, &
    \Pd_{\ga\gad}&=-\frac{i}{4}\{q_\ga,\brq_\gad\}\,, &
    \brLd_{\gad\gbd}&=-\frac{i}{4}\{\brq_\gad,\brq_\gbd\}   \,.
\end{align}
The first place where the commutators deform is
\begin{align}
    [\Pd_{\ga\gad},\Pd_{\gb\gbd}]&=(1+\mu k) \epsilon_{\ga\gb}\brLd_{\gad\gbd}+(1+\bar\mu \brk) \epsilon_{\gad\gbd}\Ld_{\ga\gb}\,.
\end{align}
This is consistent with the structure of the dual cycles \eqref{cycleLPP3d}.
The deformation that is isomorphic to the one obtained by setting $d=3$ in the free boson case corresponds to $\theta=0$ and projection by $(1+k \brk)/2$. Now, the question of uniqueness of the deformation described above is of physical significance since the resulting $A_\infty$-algebra is supposed to describe the slightly broken higher spin symmetry realized in the Chern--Simons matter theories. The double $D$ is a particular case of the smash-product algebras. The Hochschild cohomology of such algebras is known \cite{AFLS} and in our case the second Hochschild cohomology group is two-dimensional \cite{Sharapov:2017yde,Sharapov:2017lxr}. Therefore, we exhaust all possible deformations. The relevance of these statements is also discussed below.

\section{Concluding Remarks}
\label{sec:Conclusions}
The slightly broken higher spin symmetry is expected to fix all correlation functions in Chern--Simons matter theories, at least in the large-$N$ limit, which would also explain and prove the three-dimensional bosonization duality. The problem is that it is not a symmetry in any usual sense. In the free limit each CFT leads to a well-defined infinite-dimensional associative algebra, a higher spin algebra (HSA). HSA determines the correlation functions of higher spin currents. When interactions are turned on, the conservation of the higher spin currents is broken by the double-trace operators built of the currents themselves. The fact that higher spin currents, as a representation, are formally isomorphic to the HSA (up to the $\mathbb{Z}_2$-automorphism given by the inversion map $\Imap$) makes it clear that slightly broken higher spin symmetry is a deformation of a given HSA by an element of HSA itself (up to $\Imap$). While there is no place for such a deformation in the realm of associative algebras, this can be achieved by going to the $A_\infty$-setting, which is the main proposal of the paper.

One of the main results of the paper is the explicit construction of a class of $A_\infty$-algebras that can be viewed as noncommutative deformation quantization of a given associative algebra $\ass$. We show that if $\ass$ admits a deformation as an associative algebra, then we can replace the formal deformation parameter $\hbar$ by an element of $\ass$ itself by going to the $A_\infty$ setting. It turns out that the structure maps $m_n$ of the $A_\infty$-algebra are completely determined (up to a natural equivalence) by the deformation of $\ass$. Therefore, this new class of $A_\infty$-algebras is completely determined by deformations of associative algebras.

Combining these two findings, the problem of the slightly broken higher spin symmetry gets reduced to a much simpler problem of deforming the $\mathbb{Z}_2$-extension of a given HSA. This deformation may depend on several parameters, the number being given by the size of the Hochschild cohomology. We argue that there is an at least one-parameter deformation, the parameter being $1/N$. 

The case of three dimensions is special. The deformation is found to involve the pair of free parameters $\mu$ and $\bar\mu$, while it has only one deformation parameter for the free boson HSA in $d>3$. Taking the reality conditions into account, one can put $\mu=\nu e^{i\theta},\bar{\mu}=\nu e^{-i\theta}$. The microscopical description of these $3d$ CFT's with slightly broken higher spin symmetry is via the Chern--Simons matter theories with the two parameters $N$ and $k$ (in the simplest situation). So far the deformation parameters $\theta$ and $\nu$ are just phenomenological. At least in the large-$N$ limit it is possible \cite{Aharony:2012nh,GurAri:2012is} to relate them to the microscopical parameters $\theta= \tfrac{\pi}2\tfrac{N}{k}$, $\nu\sim \tilde{N}^{-1}$, $\tilde{N}=2N\tfrac{\sin \pi \lambda}{\pi \lambda}$. 
It is remarkable that the higher spin symmetry breaking in these theories is fully described by a (two copies) rather simple associative algebra of fuzzy super-sphere $S^{2|2}$. For the free case the correlation functions of higher spin currents are given by a unique invariant of the HSA --- the trace \cite{Colombo:2012jx,Didenko:2012tv,Didenko:2013bj, Bonezzi:2017vha}. Since the deformed HSA does also admit a trace, it is natural to conjecture that the correlation functions of the single-trace operators in the Chern--Simons matter theories are expressible in terms of the same type of invariants:
\begin{align}
    \langle J_1\ldots J_n\rangle && \longleftrightarrow && \mathrm{Tr}(C_1\ast \ldots \ast C_n) +\text{permutations}.
\end{align}
Here the trace is the invariant trace of the deformed higher spin algebra and $C_i$ are `wave-functions' similar to those of \cite{Colombo:2012jx,Didenko:2012tv,Didenko:2013bj, Bonezzi:2017vha}. The expression is manifestly invariant under the infinite-dimensional deformed symmetries. The dependence on $\theta$ enters implicitly via $\mathrm{Tr}$, $\ast$ and $C_i$. This is a smooth deformation of the free CFT correlation functions.

As a side remark, let us point out a relation between the deformation of the $\mathbb{Z}_2$-extended higher spin algebras (HSA) and deformation quantization. HSA result from  quantization of the algebra of functions on a Poisson manifold $M$ which is the closure of a nilpotent coadjoint orbit of $so(d,2)$, see \cite{Fronsdal2009}. A Poisson manifold may have some discrete symmetries $G$, e.g. $G=\mathbb{Z}_2$ related to the inversion map. Given $G$ there are two complementary algebras: functions on the Poisson orbifold $M/G$ and the smashed product of functions on $M$ with $G$ (we called it a double, $D(hs)$). These two algebras can have new deformations compared to the usual deformation quantization, see e.g. \cite{HOT}. The deformed HSA's we constructed are examples of this situation. The algebras depend on (at least) two deformation parameters, where the first deformation parameter (it was implicit in the paper) comes from the usual deformation quantization and the second one is due to the $\mathbb{Z}_2$-automorphism. These remarks make deformed HSA's be a part of deformation quantization.

Lastly, it is worth mentioning that the same $A_\infty$-algebras that we constructed in the paper allows one to solve \cite{Sharapov:2019vyd} the problem of Formal Higher Spin Gravities, which can indirectly, via AdS/CFT, explain the relevance of these $A_\infty$-algebras.

\section*{Acknowledgments}
\label{sec:Aknowledgements}
We are grateful to Xavier Bekaert, Maxim Grigoriev, Murat G{\"u}naydin, Carlo Iazeolla, Karapet Mkrtchyan, Dmitry Ponomarev and Ergin Sezgin for useful discussions. The work of A. Sh. was supported in part by RFBR
Grant No. 16-02-00284 A and by Grant No. 8.1.07.2018 from ``The Tomsk State University competitiveness improvement programme''. 
The work of E. S. was supported by the Russian Science Foundation grant 18-72-10123 in association with the Lebedev Physical Institute.

\begin{appendix}
\renewcommand{\thesection}{\Alph{section}}
\renewcommand{\theequation}{\Alph{section}.\arabic{equation}}
\setcounter{equation}{0}\setcounter{section}{0}

\section{First Order Deformation}
\label{app:FirstOrder}
\setcounter{equation}{0}
The first-order deformation of an $A_{-1}$-bimodule $A_0$, regarded as an $A_\infty$-algebra, is a collection of six tri-linear maps $m_3(\bullet,\bullet,\bullet)$ obeying the equations
\besubeqs
\begin{align}\label{mostimp}
    -am_3(b,c,u)+m_3(ab,c,u)-m_3(a,bc,u)+m_3(a,b,cu)&=0\,,\\
    m_3(a,b,u)c- am_3(b,u,c) +m_3(ab,u,c)-m_3(a,bu,c)-m_3(a,b,uc)&=0\,,\\
    m_3(a,u,b)c-am_3(u,b,c)+m_3(au,b,c)+m_3(a,ub,c)-m_3(a,u,bc)&=0\,,\\
    m_3(u,a,b)c-m_3(ua,b,c)+m_3(u,ab,c)-m_3(u,a,bc)&=0\,,
\end{align}
and
\begin{align}
    m_3(a,b,u)v-am_3(b,u,v)+m_3(ab,u,v)-m_3(a,bu,v)&=0\,,\\
    m_3(u,v,a)b-um_3(v,a,b)+m_3(u,va,b)+m_3(u,v,ab)&=0\,,\\
    m_3(a,u,b)v-am_3(u,b,v)+m_3(au,b,v)+m_3(a,ub,v)-m_3(a,u,bv)&=0\,,\\
    -m_3(u,a,v)b-um_3(a,v,b)-m_3(ua,v,b)+m_3(u,av,b)+m_3(u,a,vb)&=0\,,\\
    -m_3(a,u,v)b-am_3(u,v,b)+m_3(au,v,b)+m_3(a,u,vb)&=0\,,\\
    m_3(u,a,b)v-um_3(a,b,v)-m_3(ua,b,v)+m_3(u,ab,v)-m_3(u,a,bv)&=0\,,
\end{align}
\esubeqs
where $a,b,c$ are elements of $A_{-1}$ and $u,v,w\in A_0$. It is easy to see that \eqref{formA} and \eqref{formA} are solutions. These two solutions are equivalent via an $A_\infty$ change of variables, which at this order is $m_3\rightarrow m_3+\delta f$, for $f(a,u)=\phi_1(a,u)$. 

More generally, the first equation \eqref{mostimp} seems to be the most important one. Its nontrivial solutions correspond to the second Hochschild cohomology group $HH^2 (A,\mathcal{M})$, where $\mathcal{M}$ is $\mathrm{Hom}(M,M)$ endowed with the natural bimodule structure (in our case $M\sim A_0$). If $A_{-1}\sim A_1$ is the polynomial Weyl algebra on two generators, then $HH^3(A_1, N)=0$ for any bimodule $N$ as the enveloping algebra $A_1^e$ admits a projective resolution of length 2. This means that the deformations are unobstructed. The same holds true for the matrix algebras $\mathrm{Mat}_n(A_1)$ acting on the bimodule $\mathrm{Mat}_n(A_1)$ (the algebras $A_1$ and $\mathrm{Mat}_n(A_1)$ are Morita equivalent).

\section{Deformations of Higher Spin Algebras}
\label{sec:HSAdeformed}
\setcounter{equation}{0}
In order to construct the $A_\infty$-algebra description of the slightly broken higher spin symmetry we need to construct a deformation of the double $D(hs)$ of a given HSA. It is hard to prove that such deformation always exist due to the flexibility of what HSA means. We adopt Definition 3 via quotients of $U(so(d,2))$. We show that the $\Imap$-stable subalgebra, $\Imap(a)=a$, $a\in hs$, of $hs$, turns out to be the HSA of the generalized free field in $d-1$ dimension and its weight is generic. Therefore, the subalgebra admits a deformation. It turns out that this deformation can be uplifted to $D(hs)$.

\paragraph{Higher Spin Lorentz Subalgebra.} The most convenient definition of HSA at the moment  is via the universal enveloping algebra $U(so(d,2))$. Suppose we are given some $hs$ as $hs=U(so(d,2))/\Jos$ for some $\Jos$. Let us also  assume that $hs$ corresponds to some free on-shell field. The $so(d,2)$-generators $T^{AB}$ can be split into the AdS-Lorentz generators $\Ld^{\aA\aB}$ and AdS-translations $\Pd^\aA$.\footnote{For example, $\Ld^{\aA\aB}=T^{\aA\aB}$ and $\Pd^\aA=T^{\aA5}$ where $5$ is the extra dimension of an $so(d,2)$ vector as compared to an $so(d,1)$ one $\eta^{55}=-1$. Here $\aA,\aB,\ldots =0,\ldots,d$ are the indices of the AdS Lorentz algebra $so(d,1)$.} The AdS-Lorentz subalgebra $L(hs)$ of $hs$ is defined as the enveloping algebra of the $so(d,1)$ subalgebra generated by $\Ld^{\aA\aB}$. This is the stability algebra of the inversion map.\footnote{Another reason for the relevance of the AdS-Lorentz interpretation is that $\Imap \Ld^{\aA\aB}\Imap=\Ld^{\aA\aB}$ and $\Imap \Pd^\aA \Imap=-\Pd^\aA$ if we define $\Imap P^a\Imap=-K^a$ etc. Such automorphism of the AdS-algebra is used in the study of higher spin fields in AdS, see e.g. \cite{Vasiliev:2003ev}. If we define $\Imap P^a\Imap=K^a$ etc., then the stability algebra is $so(d-1,2)$, which can be interpreted as the conformal algebra in $(d-1)$-dimension. Such definition is more physical since it is the $so(d-1,2)$ subalgebra that would admit supersymmetric extensions once $so(d,2)$ does for lower $d$. Nevertheless, below we mostly use the $so(d,1)$-interpretation.}

The Lorentz subalgebra $L(hs)$ can be understood as a HSA itself ($so(d,1)$ is viewed here as the Euclidian conformal algebra in $d-1$ dimensions): it has more or less the same properties, but the Casimir value corresponds to an off-shell conformal field in $(d-1)$ dimensions.

For example, the ideal that is responsible for the free boson HSA, when $T^{AB}$ is decomposed into $\Ld^{\aA\aB}$ and $\Pd^\aA$, reads:
\besubeqs
\begin{align}
    \Jos^{\aA\aB\aC\aD}&=\Ld^{[\aA\aB} \Ld^{\aC\aD]}\,,\\
    \Jos^{\aA\aB\aC5}&=\{\Ld^{[\aA\aB}, \Pd^{\aC]}\}\,,\\
    \Jos^{\aA\aB}&= \Ld\fud{\aA}{\aC} \Ld^{\aB\aC}+\Ld\fud{\aB}{\aC} \Ld^{\aA\aC}- \Pd^\aA \Pd^\aB- \Pd^\aB \Pd^\aA-(d-2)\eta^{\aA\aB}\,,\\
    \Jos^{\aA5}&= \{ \Ld\fud{\aA}{\aC}, \Pd^{\aC}\}\,,\\
    \Jos^{55}&= 2 \Pd_\aA \Pd^\aA +(d-2)\label{PPonshell}\,,\\
    \Jos&= -\frac12 \Ld_{\aA\aB} \Ld^{\aA\aB}+\Pd_\aA \Pd^\aA+\frac14(d^2-4)\,,
\end{align}
\esubeqs
from which it follows
\begin{align}
    \Jos^{\aA\aB\aC\aD}&=\Ld^{[\aA\aB} \Ld^{\aC\aD]}\,, &
    \Jos&= -\frac12 \Ld_{\aA\aB} \Ld^{\aA\aB}+\frac{d}{4}(d-2)\,. \label{CasimirLor}
\end{align}
This is exactly the ideal that defines the HSA of the generalized free field, but in one dimension lower, cf. \eqref{JGFF} and \cite{Joung:2015jza}. The conformal weight of this fictitious generalized free field in $(d-1)$ dimensions is $(d-2)/2$ or $d/2$.\footnote{The value of the Casimir operator is $\Delta(\Delta-(d-1))$. Notice that both the roots are above the unitarity bound $(d-1)/2-1$. } 

Thanks to the fact that the weight of this fictitious generalized free field is generic the Lorentz subalgebra belongs to a one-parameter family of algebras. Therefore, the Lorentz subalgebra can be deformed. In particular, the second Hochschild cohomology group is not empty, $HH^2(L(hs),L(hs))\neq 0$.

\paragraph{Deformation of the Double.} That the $\Imap$-stable subalgebra $L(hs)$ admits a one-parameter family of deformations is an indication that the $\Imap $-extended algebra $D(hs)$ also does. It seems to be no general theorem, however, that would allow one to directly construct such a deformation.\footnote{Indeed, the same argument applies to the $hs$ itself, while it is usually rigid, as different from $D(hs)$. } The following three justifications are helpful. (1) In the case of the smash-product of the Weyl algebra by a finite group of symplectic reflections (which is the case that many HSA's can be reduced to) it can be shown that such deformations do exist and it is even possible to explicitly construct them, see \cite{Sharapov:2017lxr,Sharapov:2018hnl}. Note that the Weyl algebra itself is rigid and therefore extending it with $\mathbb{Z}_2$ is crucial for the deformation. (2) For many algebras there is a duality \cite{Bergh} between Hochschild homology and cohomology and we can explicitly construct the cycle that the sought for Hochschild two-cocycle is dual to (see below). (3) At least for the algebras we are interested in this paper there is a simple oscillator realization and in Section \ref{sec:Oscillators} we construct the deformed double $D_\hbar(hs)$ explicitly.

\paragraph{Dual Cycle.} Cochains act in a natural way on  chains, so that the latter form a module over the former \cite{Shoikhet:2000gw}. As different from cocycles, cycles are usually easier to find. Then, if the algebra falls into the class of algebras for which the Hochschild cohomology $HH^\bullet(A)$ is dual to the homology $HH_\bullet(A)$, one can compute the dimension of various $HH^\bullet(A)$ from those of $HH_\bullet(A)$. Another usage of nontrivial cycles is to test whether a given cocycle is nontrivial since the chain differential is dual to the cochain differential with respect to the natural pairing. We will construct a cycle for $D(hs)$, which implies that there is a dual cocycle.

Note first, that the HSA $hs$ of some free field\footnote{Here we avoid generalized free fields at generic value of the conformal weight.} is determined by a certain two-sided ideal $\Jos$ of $U(so(d,2))$. For the free fields obeying the $\square$-type equation the ideal contains the generator described by the Young diagram $\YoungB$.  Taken together with the fixed value of the Casimir operator this means that the $AdS$-momentum squares to a constant:
\begin{equation}\label{pp}
    \Pd_\aA \Pd^\aA=M^2\,.
\end{equation}
For example, for the HSA of the free boson CFT we find \cite{Iazeolla:2008ix} \eqref{PPonshell}\footnote{This is the AdS-base rewriting of $P^aP_a=0$, $K^aK_a=0$, and $\boldsymbol{C}_2-\tfrac14(d^2-4)=0$.}
\begin{equation}\label{ppFB}
    \Pd_\aA \Pd^\aA=-\frac{(d-2)}{2}\,.
\end{equation}
Now, consider the two-chain\footnote{The Hochschild differential acts as (note the twist by $\Imap$) $$\pl (c_0\otimes c_1\otimes\cdots \otimes c_k )=c_0c_1\otimes c_2\cdots \otimes c_k-c_0\otimes c_1c_2\otimes\cdots \otimes c_k+\cdots +(-)^k I(c_k) c_0\otimes c_1\cdots \otimes c_{k-1}\,.$$ The arguments $c_1,\ldots, c_k$ are assumed to take values in the quotient space $hs/K$, where $K\subset hs$ is the base field. In practice this means that $K\sim 0$ for all the factors except for the first one. Such complex is called {\it normalized} and it is known to have the same homology, $HH_\bullet(hs/K, hs^\Imap)\sim HH_\bullet(hs,hs^\Imap)$. }
\begin{equation}
    \gamma= 1\otimes \Pd_\aA\otimes \Pd^\aA \,.
\end{equation}
It is a nontrivial cycle of $hs$ with values in the representation $hs^I$ that is twisted by $\Imap $:
\begin{equation}
   \partial \gamma =\Pd_\aA\otimes \Pd^\aA-1\otimes \Pd_\aA \Pd^\aA+\Imap(\Pd^\aA)\otimes \Pd_\aA=0\,.
\end{equation}
Here we used \eqref{pp} and the fact that the complex is normalized, i.e., $M^2\sim 0$ when it appears in any of the factors except the first one. In this case it is easy to uplift the cycle from the normalized complex to the original one. Indeed, 
\begin{equation}
    \gamma'= 1\otimes \Pd_\aA\otimes \Pd^\aA+M^2 (1\otimes 1\otimes 1) 
\end{equation}
is closed as it is. Therefore, $\gamma'$ represents a class in $HH_2(hs,hs^I)$. This cycle can also be uplifted to the cycle of the full double $D(hs)$
\begin{equation}\label{gamma'}
    \gamma'= \Imap\otimes \Pd_\aA\otimes \Pd^\aA+M^2 (\Imap \otimes 1\otimes 1) \,, 
\end{equation}
representing thus an element of $HH_2(D(hs),D(hs))$. It should be dual to a nontrivial cocycle $\phi$ representing an element of $HH^2(D(hs),D(hs))$. This is the cocycle \eqref{Ideform} we need to deform $D(hs)$.

Let us also consider the special case of three dimensions. Firstly, we can replace (\ref{gamma'}) with an equivalent two-cycle
\begin{align}\label{cycleLPP}
    \gamma'&=\Ld_{\aA\aB}\otimes \Pd^\aA\otimes \Pd^\aB+\frac14 1\otimes \Ld_{\aA\aB}\otimes \Ld^{\aA\aB}-\frac12 C_L 1\otimes 1\otimes 1\,, &&\pl \gamma'=0\,,
\end{align}
where $C_L=-\tfrac12 \Ld_{\aA\aB} \Ld^{\aA\aB}$ is the value of the Casimir operator of the AdS-Lorentz subalgebra, see e.g. \eqref{CasimirLor}. Secondly, in the $sl(2,\mathbb{C})$-spinorial language the generators $T^{AB}$ of $so(3,2)$ decompose into $\Pd_{\ga\gad}$, and $\Ld_{\ga\gb}$, $\Ld_{\gad\gbd}$,\footnote{Here, $\ga,\gb$ and $\gad,\gbd$ are the indices of the fundamental representation of $sl(2,\mathbb{C})$ and its conjugate. The dictionary between the vectorial and spinorial languages is via the $\sigma$-matrices, e.g. $\Pd_\aA=\sigma^{\ga\gad}_\aA \Pd_{\ga\gad}$. } the latter being (anti)-selfdual components of $\Ld_{\aA\aB}$. Then, \eqref{cycleLPP} reduces to the two independent cycles:
\begin{align}\label{cycleLPP3d}
    \gamma'&=\Ld^{\ga\gb}\otimes \Pd_{\ga\dot\beta}\otimes \Pd\fdu{\gb}{\dot\beta}+\frac12 1\otimes \Ld_{\ga\gb}\otimes \Ld^{\ga\gb}- c_L 1\otimes 1\otimes 1\,, &&\pl \gamma'=0\,,
\end{align}
where $c_L=-\frac12 \Ld_{\ga\gb}\Ld^{\ga\gb}=-3/4$ and the second cycle is obtained by complex conjugation. These two cycles imply that there are two independent deformations of the free boson HSA and free fermion HSA (which is the same) in three dimensions. In Section \ref{sec:Oscillators} we provide a full description of the deformed algebra for the examples of interest. In order to better understand the reason for the extended higher spin algebras to admit a deformation it would be instructive to study the representation theory of the deformed algebras and its field-theoretical realizations.

\section{Sketch of the Proof}
\label{app:Proof}
\setcounter{equation}{0}
We need to check that $m_n$ defined in Section \ref{sec:Aconstruction} do solve the Maurer--Cartan equation
\begin{align}\label{MCapp}
    \delta m_n +\sum_{i+j=n+2} m_i \circ m_j&=0\,.
\end{align}
Due to the specific form of $m_n$ (with arguments from $A_{-1}$ on the left) there are fewer equations to be checked. Firstly, one can restrict oneself to the sector with three $A_{-1}$ factors and $n-2$ factors in $A_0$, i.e., the arguments in \eqref{MCapp} are permutations of $abcuv\ldots w$. Secondly, the nontrivial equations can be parameterized by the position of $c$:   
\begin{align}\label{MCappA}
    E_k(a,b,\ldots ,u,c,\overbrace{v,\ldots ,w}^k)&=\delta m_n +\sum_{i+j=n+2} m_i \circ m_j\Big|_{a,b,\ldots ,u,c,v,\ldots,w}=0\,.
\end{align}
The differential $\delta$ is very simple for most of $k$'s, $k=1,\ldots ,n-3$:
\begin{align}\label{diffA}
    \delta m_n(a,b,\ldots ,u,c,v,\ldots ,w)=-m_n(a,b,\ldots ,uc,v,\ldots ,w)+m_n(a,b,\ldots,u,cv,\ldots,w)
\end{align}
and contains four terms for the maximal $k=n-2$
\begin{align}\label{diffB}
\begin{aligned}
    \delta m_n(a,b,c,v,\ldots ,w)&=-am_n(b,c,v,\ldots,w)+m_n(ab,c,v,\ldots,w)+\\
    &\qquad-m_n(a,bc,v,\ldots,w)+m_n(a,b,cv,\ldots,w)\,.
\end{aligned}
\end{align}
The rationale for the recursive formula given in the main text is that the differential \eqref{diffA} annihilates those components of $m_n$ that have too many multiplicative arguments on the right. Therefore, one can start at $k=1$, to which only $m_3$ and $m_{n-1}$ contribute:
\begin{align*}
    E_1(a,b,\ldots ,u,c,w)=& -m_n(a,b,\ldots,uc,w)+m_n(a,b,\ldots,u,cv)+\\
    &-m_{n-1}(a,b,\ldots,m_3(u,c,w))+m_3(m_{n-1}(a,b,\ldots,u),c,v)=0\,.
\end{align*}
This equation determines the part of $f_n$ that has no multiplicative arguments at all. Using $m_{n-1}=f_{n-1}(a,b,\ldots)u$ and explicit form of $m_3$, one observes that $f_n=\phi_1(f_{n-1}(a,b,\ldots),u)$, i.e., $m_n=f_n(a,b,\ldots,u)w$, up to the terms with more direct factors. Next, one should proceed to $k=2$ and fix the part of $f_n$ that has one direct factor. At each order one will get the equations that are supposed to be true for $m_{3,\ldots,k+2}$. The trick here is that $E_k$ contains Gerstenhaber products of $m_3, m_{n-1}$, \ldots, $m_{k+2}, m_{n-k}$ and the lowest $f_{n-k}$ always enters with the same arguments, i.e., can be treated as a single variable. Eventually, $E_k$ can be reduced to equations for $m_3$,\ldots, $m_{k+2}$ irrespective of $n$. For example, $E_1=0$ is solved by $f_n=\phi_1(f_{n-1},\bullet)$ irrespective of what $f_{n-1}$ is, but the same time this fixes the lowest component of $f_{n-1}$ itself, and so on.

A non-recursive proof is based on manipulations with the trees. Let us recall that $f_n$ is a sum over all terms that are depicted by trees (with one branch)\footnote{Recall that the (green) dots correspond to some $\phi_{m+1}$, while the simple vertices are mapped into insertions of multiplicative arguments on the right.}
\begin{align}\label{treeOneB}
f_n(a,b,u,\ldots,w)&\ni \quad
\parbox{4cm}{\begin{tikzpicture}[scale=0.4]
    \def\R{0.15}
    \draw[black,thick] node[below]{$\scriptstyle a$} (0,0) -- (3.5,3.5);
    \draw[black,thick] (1,1) -- (2,0) node[below]{$\scriptstyle b$};
    \filldraw[color=black,fill=green]   (1,1)  circle (\R);
    \draw[black,thick] (1.5,1.5) -- (2,1.0) node[below right]{$\scriptstyle u$};
    \draw[black,thick] (2,2) -- (2.5,1.5);
    \draw[black,thick] (2.5,2.5) -- (3,2);
    \draw[black,thick] (3,3) -- (4,2);
    \filldraw[color=black,fill=green]   (3,3)  circle (\R);    
    \draw[black,thick] (3.9,3.9) -- (6.5,6.5);
    \draw[black,thick] (4.5,4.5) -- (5,4);    
    \draw[black,thick] (5,5) -- (6,4);
    \filldraw[color=black,fill=green]   (5,5)  circle (\R);    
    \draw[black,thick] (5.5,5.5) -- (6,5);
    \draw[black,thick] (6,6) -- (6.5,5.5);    
    \draw[black,thick] (6.5,6.5) -- (7,6) node[below right]{$\scriptstyle w$};
    \end{tikzpicture}}
\end{align}
To deal with more complicated trees we introduce an order. A tree is called ordered if it does not contain vertices of the form
\begin{align}
    \orderBA{m+1}{0}
\end{align}
i.e., in the actual expression any $\phi_{m+1}(\bullet,\bullet)$ does not have any factors on the left, e.g. $a\phi_{m+1}(b,c)$, where $a$ can be any expression possibly containing several factors and other $\phi$'s. The bad vertices can be ordered via 
\begin{align}
    \parbox{1.6cm}{\orderBA{k}{0}}=\parbox{1.6cm}{\orderB{0}{k}}-\parbox{1.6cm}{\orderBA{0}{k}}+\parbox{1.6cm}{\orderB{k}{0}}+\sum_{i+j=k}\parbox{1.6cm}{\orderB{j}{i}}-\parbox{1.6cm}{\orderBA{j}{i}}
\end{align}
Equation $E_k$ contains several terms, those coming from $\delta m_n$ are already ordered (except for $E_{n-2}$). Also, only good vertices arise when $m_i$ is inserted into $m_j$ as an (right) argument of some $\phi_k$:
\begin{align}\label{treeTwoB}
\parbox{4cm}{\begin{tikzpicture}[scale=0.4]
    \def\R{0.15}
    \draw[black,thick] (0,0) -- (3.5,3.5);
    \draw[black,thick] (1,1) -- (2,0);
    \filldraw[color=black,fill=green]   (1,1)  circle (\R);
    \draw[black,thick] (1.5,1.5) -- (2,1.0);
    \draw[black,thick] (2,2) -- (2.5,1.5);
    \draw[black,thick] (2.5,2.5) -- (3,2);
    \draw[black,thick] (3,3) -- (4,2);
    \filldraw[color=black,fill=green]   (3,3)  circle (\R);    
    \draw[black,thick] (3.9,3.9) -- (6.5,6.5);
    \draw[black,thick] (4.5,4.5) -- (5,4);    
    \draw[black,thick] (5,5) -- (7,3);
    \filldraw[color=black,fill=green]   (5,5)  circle (\R);    
    \draw[black,thick] (5.5,5.5) -- (6,5);
    \draw[black,thick] (6,6) -- (6.5,5.5);    
    \draw[black,thick] (6.5,6.5) -- (7,6) ;
    \draw[black,thick] (6.5,3.5) -- (3,0);
    \draw[black,thick] (4,1) -- (5,0);
    \filldraw[color=black,fill=green]   (4,1)  circle (\R);
    \draw[black,thick] (4.5,1.5) -- (5,1);
    \draw[black,thick] (5,2) -- (5.5,1.5);
    \draw[black,thick] (5.5,2.5) -- (6.5,1.5);
    \filldraw[color=black,fill=green]   (5.5,2.5)  circle (\R);
    \draw[black,thick] (6,3) -- (6.5,2.5);
    \end{tikzpicture}}
\end{align}
The only source of bad vertices is when $m_i$ in inserted into an argument of $m_j$ that corresponds to a multiplicative argument (simple vertex):
\begin{align}
\parbox{4cm}{\begin{tikzpicture}[scale=0.4]
    \def\R{0.15}
    \draw[black,thick] (0,0) -- (3.5,3.5);
    \draw[black,thick] (1,1) -- (2,0);
    \filldraw[color=black,fill=green]   (1,1)  circle (\R);
    \draw[black,thick] (1.5,1.5) -- (2,1.0);
    \draw[black,thick] (2,2) -- (2.5,1.5);
    \draw[black,thick] (2.5,2.5) -- (3,2);
    \draw[black,thick] (3,3) -- (4,2);
    \filldraw[color=black,fill=green]   (3,3)  circle (\R);    
    \draw[black,thick] (3.9,3.9) -- (6.5,6.5);
    \draw[black,thick] (4.5,4.5) -- (6.5,2.5);    
    \draw[black,thick] (5,5) -- (6,4);
    \filldraw[color=black,fill=green]   (5,5)  circle (\R);    
    \draw[black,thick] (5.5,5.5) -- (6,5);
    \draw[black,thick] (6,6) -- (6.5,5.5);    
    \draw[black,thick] (6.5,6.5) -- (7,6) ;
    \draw[black,thick] (6,3) -- (3,0);
    \draw[black,thick] (4,1) -- (5,0);
    \filldraw[color=black,fill=green]   (4,1)  circle (\R);
    \draw[black,thick] (4.5,1.5) -- (5,1);
    \draw[black,thick] (5,2) -- (5.5,1.5);
    \draw[black,thick] (5.5,2.5) -- (6.5,1.5);
    \filldraw[color=black,fill=green]   (5.5,2.5)  circle (\R);
    \end{tikzpicture}}
\end{align}
These terms need to be reordered and will eventually generate (with the opposite sign) all the trees with one branch \eqref{treeOneB} or two branches \eqref{treeTwoB} that are already present. Therefore, the Maurer--Cartan equation is indeed satisfied. It would be interesting to find an appropriate configuration space where the proof would reduce to the Stokes theorem.

\end{appendix}
\providecommand{\href}[2]{#2}\begingroup\raggedright\endgroup

\end{document}